%% file: main.tex
\documentclass[journal]{IEEEtran}
\usepackage{graphicx}
\usepackage{amsmath}
\usepackage[colorinlistoftodos]{todonotes}
\usepackage{subfigure}
\usepackage{multirow}
\usepackage[colorlinks,linkcolor=blue]{hyperref}
\usepackage{pdfpages}
\usepackage{url}
\usepackage{amsfonts}
\usepackage{booktabs}
\usepackage{pifont}
\usepackage{cleveref}
\usepackage[ruled,vlined,linesnumbered]{algorithm2e}
\SetKwInput{KwInput}{Input}
\SetKwInput{KwOutput}{Output}
\SetKwComment{Comment}{$\triangleright$\ }{}
\DontPrintSemicolon

\newcommand{\gain}[1]{$_{\textcolor{red}{#1}}$}
\newcommand{\same}[1]{$_{\textcolor{gray}{#1}}$}

\begin{document}
\title{{\fontsize{23pt}{20pt}\selectfont
Cross-Modal Ultrasound-MRI Learning for Fetal Brain Ventricular Volumetry and Abnormality Screening}}
\author{Yuhao Huang, Yuanji Zhang, Yuhuan Lu, Dong Ni, P. Ellen Grant, Davood Karimi
\thanks{Yuhao Huang, Yuanji Zhang, Yuhuan Lu, P. Ellen Grant, and Davood Karimi are with Department of Radiology, Boston Children’s Hospital, and Harvard Medical School, Boston, USA. (Corresponding author: Davood Karimi, email: Davood.Karimi@childrens.harvard.edu)}
\thanks{Yuhao Huang, Yuanji Zhang, and Dong Ni are with Medical Ultrasound Image Computing (MUSIC) Lab, Shenzhen University, Shenzhen, China}
\thanks{Yuhao Huang is also with Centre for Artificial Intelligence and Robotics, Hong Kong Institute of Science \& Innovation, Chinese Academy of Sciences, Hong Kong, China. Yuanji Zhang is also with Shenzhen Luohu People's Hospital, Shenzhen, China. Dong Ni is also with School of Artificial Intelligence, Shenzhen University, Shenzhen, China and School of Biomedical Engineering and Informatics, Nanjing Medical University, Nanjing, China.}
\thanks{This work was supported by the Frontier Technology Development Program of Jiangsu Province (No. BF2024078) and National Natural Science Foundation of China (No. 12326619)}
}

\maketitle

\begin{abstract}
Assessment of ventriculomegaly (VM) on fetal brain ultrasound relies primarily on measuring lateral ventricular atrial width on standard planes, which is operator-dependent and may not fully reflect the overall ventricular enlargement. Fetal brain MRI provides more reliable volumetric information but is costly and less accessible for routine use. To address these limitations, we propose VIFBA, an ultrasound video-based framework for fetal brain assessment that predicts MRI-derived lateral ventricular volume, classifies VM severity, and identifies potential non-VM fetal brain abnormalities. 
Our contribution is three-fold. 
First, we introduce a joint-embedding predictive architecture (JEPA)-inspired tube latent prediction objective that leverages spatio-temporal coherence in ultrasound videos to enhance representation learning. 
Second, we develop a contrastive cross-modal alignment strategy that transfers structural information from MRI to ultrasound during training, while requiring ultrasound alone at inference. 
Third, we augment VIFBA with a training-free vision-language model and retrieval augmentation to verify uncertain predictions and identify potential non-VM fetal brain abnormalities. 
We validated VIFBA on a large dataset comprising 857 cases (3,196 videos) with paired fetal brain ultrasound and MRI examinations. 
On held-out test data, VIFBA achieved an MAE of 0.5909 mL and Pearson correlation coefficient of 0.9907 for ventricular volume regression, 0.9400 accuracy for VM severity classification, and an F1 score of 0.7764 for multi-abnormality classification, substantially outperforming single-task baselines, video-based strong competitors, and state-of-the-art foundation models. By enabling MRI-informed volumetric assessment from routine ultrasound alone, VIFBA offers a practical and potentially broadly deployable pathway toward accurate and affordable prenatal brain screening.
\end{abstract}


\section{Introduction}
\label{sec:introduction}
\input{intro}

\section{Related work}
\label{sec:related_work}
\input{related_work}

\section{Method}
\label{sec:method}
\input{method}

\section{Experiments}
\label{sec:exp}
\input{exp}

\section{Discussion and Conclusion}
\label{sec:conclusion}
\input{conclusion}

\bibliographystyle{ieeetr}
\bibliography{ref}
\end{document}

%% file: intro.tex
Ventriculomegaly (VM), defined as the abnormal enlargement of the fetal lateral ventricles, is among the most common central nervous system abnormalities~\cite{pagani2014neurodevelopmental}. 
Its clinical significance lies in the wide variability of outcomes, ranging from normal neurodevelopment to severe impairment, depending on its severity and underlying etiology.
Therefore, accurate quantification of ventricular size, alongside reliable VM diagnosis, is critical for the detection and severity assessment of VM, as well as for risk stratification and clinical decision-making.

In prenatal ultrasound examinations, VM is primarily assessed by measuring the atrial width of the lateral ventricles on selected standard brain planes.
Based on this measurement, VM is defined as $\geq$10 mm and further stratified into mild (10–12 mm), moderate (13–15 mm), and severe ($>$15 mm) categories, while values $<$10 mm are considered normal~\cite{fox2018mild}.
However, this process requires the accurate identification of the standard plane and key anatomical landmarks to measure the maximal atrial width, making it highly dependent on operator expertise and susceptible to error accumulation~\cite{malinger2020isuog}.

Compared to 2D width measurements, ventricular volume provides a global quantitative measure that better reflects the full spectrum of ventricular conditions, from normal morphology to ventricular enlargement.
In clinical practice, volumetric assessment typically relies on 3D imaging modalities, such as 3D ultrasound and magnetic resonance imaging (MRI).
Among these, MRI generally provides clearer structural boundaries of the lateral ventricles, enabling more reliable segmentation of the ventricles for volumetric estimation~\cite{kuklisova2012reconstruction}.
However, MRI is costly and not routinely accessible, limiting its use in large-scale prenatal screening.
This motivates the development of automatic methods that enable direct estimation of ventricular volume from ultrasound scans.

Despite these motivations, this task remains highly challenging for several reasons. 
First, it relies on paired ultrasound–MRI data, which are scarce in clinical practice and difficult to acquire at scale.
Second, fetal brain ultrasound examinations typically consist of a variable number of videos acquired from different views (e.g., sagittal and axial planes), making effective multi-view information aggregation non-trivial.
Third, VM severity assessment involves fine-grained categorization (i.e., mild/moderate/severe), leading to inherent data imbalance across categories, particularly for severe cases.
Moreover, other fetal brain abnormalities may also be present, further increasing the complexity of the task.

To address these challenges, we propose a novel framework, termed VIFBA (\textbf{VI}deo-based \textbf{F}etal \textbf{B}rain \textbf{A}ssessment), for joint estimation of ventricular volume and VM severity from ultrasound videos.
VIFBA can be flexibly integrated with different ultrasound foundation models as shared backbones and employs two task-specific branches for ventricular volume regression and VM classification, respectively.
Its contribution is three-fold.
First, we introduce a tube latent prediction loss to enhance representation learning by leveraging spatio-temporal coherence in ultrasound videos.
Second, we design a contrastive cross-modal fusion strategy to inject MRI information into ultrasound representations during training, enabling MRI-informed feature learning from ultrasound alone at inference.
Third, we extend VIFBA by integrating the retrieval-augmented training-free visual-language model (VLM) to further optimize the uncertain predictions and alert the potential non-VM brain abnormality.
VIFBA was validated on a large ultrasound-MRI paired dataset with 857 cases, demonstrating its effectiveness for fetal lateral ventricular analysis.

%% file: related_work.tex
\subsection{Deep Learning in Fetal Brain Analysis}

In recent years, deep learning has driven rapid progress in prenatal imaging~\cite{baumgartner2017sononet,namburete2018fully,yang2021searching,liang2026prototype,zhang2026artificial}, particularly in automated and intelligent fetal brain analysis. 
Clinically, fetal brain assessment mainly relies on ultrasound and MRI, where ultrasound serves as the primary modality for routine prenatal screening, while fetal MRI is commonly used as a complementary tool for clearer visualization of intracranial structures and further evaluation of suspected abnormalities~\cite{malinger2020isuog,griffiths2017use}. Accordingly, existing studies have mainly evolved along two relatively distinct directions, namely ultrasound-based analysis and MRI-based analysis, which are reviewed separately below.

In fetal brain ultrasound, a major line of research has focused on automatic standard plane localization~\cite{baumgartner2017sononet,yang2021searching,yang2021agent,dou2025standard}, view classification~\cite{huang2023fourier}, and quality assessment~\cite{liu2023hierarchical,guo2024unsupervised,chen2025enhancing}, aiming to ensure diagnostically reliable views for subsequent analysis.  
Early downstream studies mainly concentrated on segmentation or measurement of single or limited anatomical targets and parameters, including segmentation of the fetal head~\cite{liu2020remove,zeng2021fetal,huang2022online}, choroid plexus and corpus callosum~\cite{huang2018learning}, and cerebellum~\cite{shu2022ecau}, as well as measurement of head circumference~\cite{zeng2021fetal,sinclair2018human}, biparietal diameter~\cite{sinclair2018human}, and lateral ventricular width~\cite{chen2020automatic}.
Recent studies have further moved toward automatic delineation and measurement of multiple structures~\cite{coronado2023automatic}, as well as fine-grained segmentation of 25 anatomical structures across five standard planes~\cite{ma2026structure}.
Xie et al.~\cite{xie2020using} investigated deep learning algorithms for classifying normal or abnormal brains. More recently, Duan et al.~\cite{duan2025fetalflex} proposed an anatomy-guided diffusion framework to synthesize normal and abnormal brain images, showing its potential to improve downstream performance.

In fetal brain MRI, existing studies have mainly focused on volumetric reconstruction, anatomical segmentation, and quantitative structural analysis.
Since fetal motion often leads to inter-slice misalignment in clinical MRI acquisitions, many methods first aim to reconstruct high-quality brain volumes from stacks of 2D slices~\cite{xu2023nesvor,firenze2026fast}.
Based on reconstructed volumes, subsequent studies have developed automated segmentation and parcellation methods for fetal brain tissues and anatomical regions, enabling quantitative analysis of various structures~\cite{payette2021automatic,uus2023bounti,ciceri2023review,zalevskyi2026advances,zeng2026atlas}.
Vahedifard et al.~\cite{vahedifard2023automatic} automated VM detection by measuring lateral ventricular width on reconstructed fetal MRI planes.
More recently, Huang et al. proposed BrainSeg~\cite{huang2026brainseg} for generalized brain tissue segmentation, parcellation, and lesion labeling across diverse MRI contrasts and populations spanning fetal to adult stages.

Despite notable progress in both ultrasound- and MRI-based fetal brain analysis, most existing studies investigate the two modalities separately and rarely exploit their complementary information. 
Moreover, VM assessment is still largely limited to 2D plane-based linear measurements, which may not fully reflect the overall volumetric enlargement of the lateral ventricles. In addition, abnormality analysis often focuses on coarse normal/abnormal classification or a few common diseases, leaving diverse and coexisting fetal brain abnormalities insufficiently explored.
Besides, to the best of our knowledge, existing studies lack cross-modal predictive investigations, particularly the task of estimating MRI-derived volumetric measurements directly from ultrasound videos.

\subsection{Medical Ultrasound Foundation Model}

Foundation models have recently become an important paradigm in medical image analysis, as large-scale pretraining enables transferable representations that can be adapted to diverse downstream tasks with limited annotations~\cite{yan2024foundation,huang2024robust,ou2026foundation}.
For example, BiomedCLIP learned transferable biomedical vision-language representations from large-scale image-text pairs~\cite{zhang2023biomedclip}.
Medical segment anything models (SAMs) demonstrated strong generalization ability for promptable image segmentation~\cite{ma2024segment,huang2024segment}.
MedSAM2 further adapted the SAM2 paradigm to both videos and 3D images, enabling promptable segmentation with spatial or temporal consistency across slices and frames~\cite{ma2025medsam2}.
Beyond SAM, several studies have also explored Self-Distillation with No Labels (DINO) variants~\cite{caron2021emerging,oquab2023dinov2} for various medical image analysis tasks~\cite{song2024general,scholz2025mm,perez2025exploring}, achieving promising performance.

Recently, foundation models have also been increasingly explored in general ultrasound imaging. 
USFM leveraged a large-scale multi-organ, multi-center, and multi-device ultrasound dataset to pretrain the foundation model, showing strong label efficiency across segmentation, classification, and enhancement tasks~\cite{jiao2024usfm}. 
EchoCare further increased the scale of ultrasound pretraining data (2 million$\rightarrow$4.5 million images), and validated its generalizability on a broader range of downstream clinical tasks~\cite{zhang2025fully}.
Most recently, Ultrasound-CLIP explored semantic-aware contrastive pretraining for general ultrasound understanding using 365k image-text paired samples~\cite{jin2026ultrasound}.

In cardiac ultrasound, EchoCLIP~\cite{christensen2024vision} and EchoPrime~\cite{vukadinovic2026comprehensive} were introduced to learn vision-language representations from echocardiography image- or video-text pairs for comprehensive cardiac evaluation.
EchoONE investigated SAM-based multi-plane echocardiography segmentation within a unified model~\cite{hu2025echoone}.
Recently, FrameONE explored the multi-view keyframe detection task in cardiac videos~\cite{chen2026frameone}.
In fetal ultrasound, FetalCLIP first learned generalizable fetal ultrasound representations from paired image-text data~\cite{maani2026fetalclip}, while Sonomate introduced a visually grounded language model by aligning ultrasound video features with transcribed sonographer speech for anatomy detection and visual question answering~\cite{guo2026visually}. 
These studies show the potential of foundation models for fetal ultrasound understanding; however, their application to fetal brain video analysis, especially ventricular volume estimation and fine-grained VM severity and brain abnormality assessment, remains insufficiently explored.

\subsection{LLM/VLM-assisted Medical Image Analysis}

Recent studies have explored the integration of large language models (LLMs) or VLMs with specialized medical imaging systems to support downstream analysis. ChatCAD~\cite{wang2023chatcad} and ChatCAD+~\cite{zhao2024chatcadplus} integrated outputs from computer-aided diagnosis models with LLMs for unified interpretation and reliable report generation. In prenatal ultrasound, FAA-Net~\cite{liang2026faa} incorporated LLM-derived medical knowledge into a multi-instance learning framework to identify diagnostically relevant information for abdominal anomaly analysis.
Most recently, See-in-Pairs~\cite{jin2025see} provided VLMs with matched reference images for comparative medical diagnosis, while RAD-SRAC~\cite{klaudel2025rad} employed retrieved examples as contextual guidance for training-free radiological image classification.

Nevertheless, existing studies mainly focus on directly supporting diagnosis, classification, report generation, or interpretation. Their applicability remains relatively limited in scenarios that require the reliability verification of existing model predictions or the identification of abnormalities not represented during training.
This limitation is particularly relevant to our complex fetal brain analysis, where heterogeneous clinical evidence, including current ultrasound videos, model predictions, and retrieved visual references and textual report information, needs to be jointly considered to assess quantitative regression and categorical outputs, while potential coexisting abnormalities also need to be identified.

%% file: method.tex
\begin{figure*}[!t]
    \centering
    \includegraphics[width=1.0\textwidth]{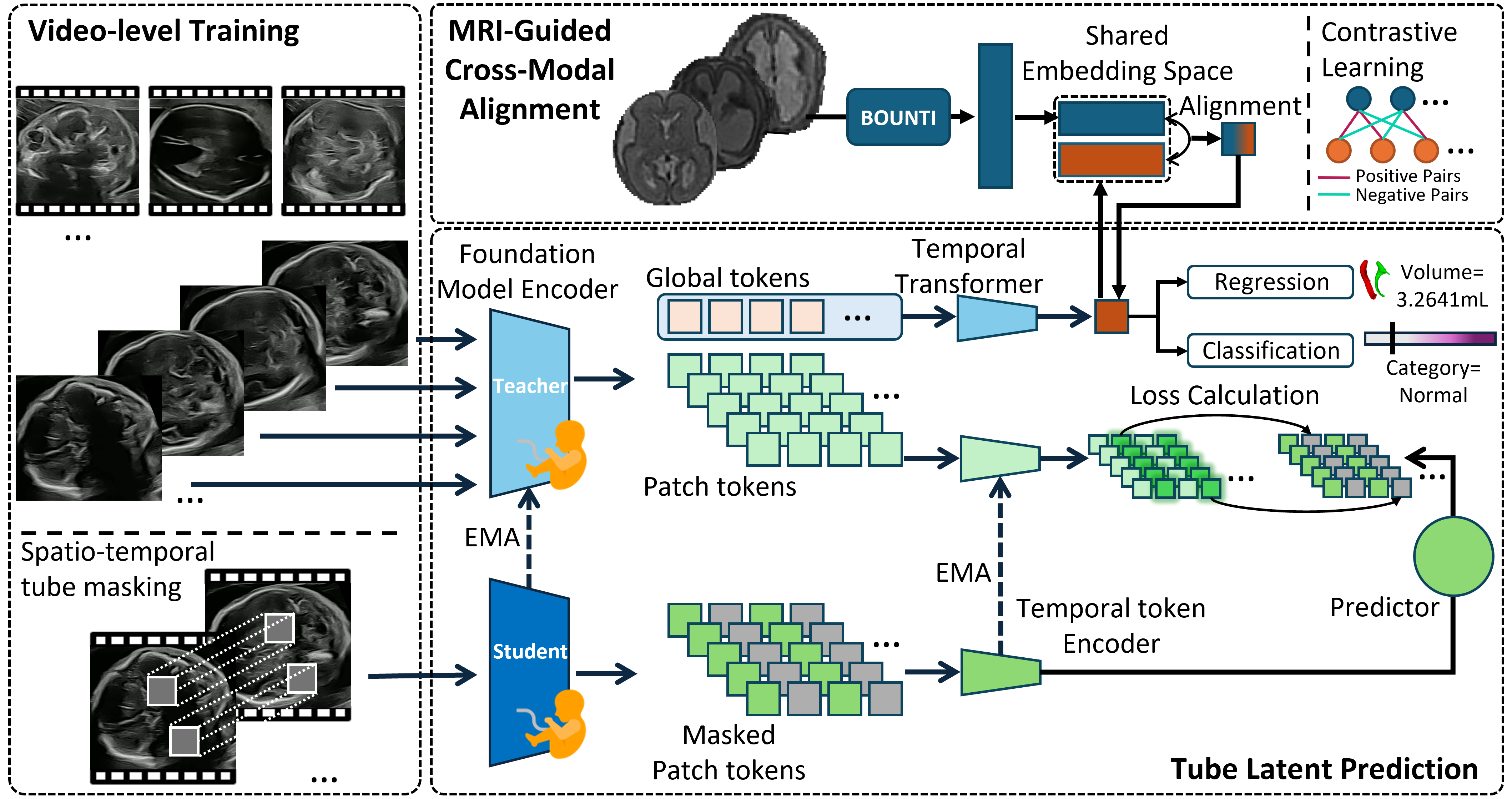}
    \caption{Overview of the training pipeline of our proposed method.} 
    \label{fig:framework}
\end{figure*}

\begin{figure*}[!t]
    \centering
    \includegraphics[width=1.0\textwidth]{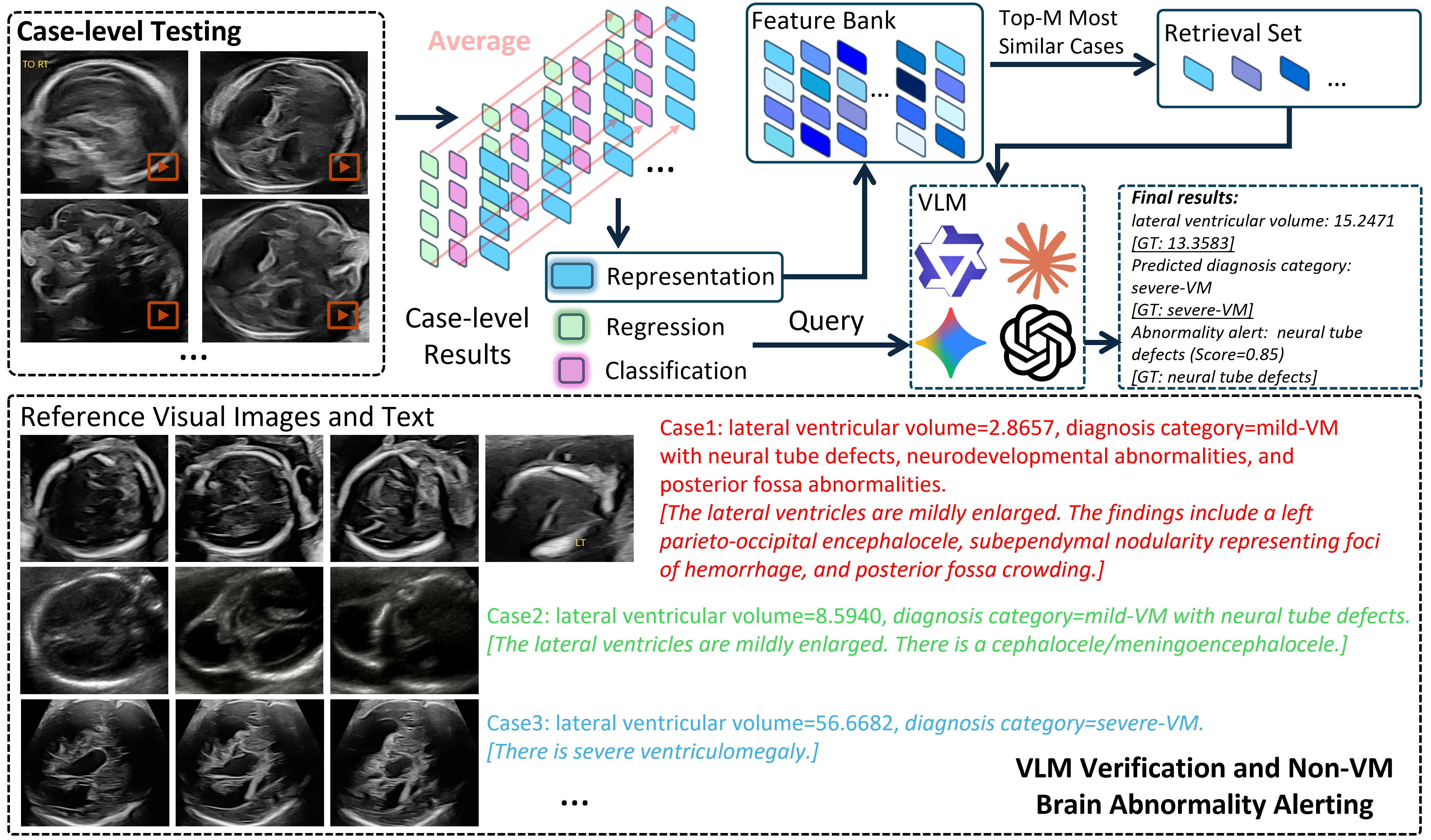}
    \caption{Testing pipeline of our proposed method, including case-level testing, and VLM verification and non-VM brain abnormality alerting.} 
    \label{fig:framework2}
\end{figure*}

Figures~\ref{fig:framework} and~\ref{fig:framework2} show the overall framework of our proposed VIFBA.
During training, each ultrasound video is treated as an individual training sample and encoded using a pre-trained vision foundation model to learn shared representations, followed by two task-specific branches for ventricular volume regression and VM classification. In addition, paired MRI data are leveraged to provide complementary supervision, enabling MRI-informed feature learning. 
During inference, predictions from multiple videos belonging to the same case are aggregated at the case level, where VIFBA relies solely on ultrasound videos and incorporates a training-free VLM with retrieval augmentation to refine uncertain predictions, making the overall framework more flexible for variable clinical acquisition protocols.
Details refer to the following sections.

\subsection{Foundation Model-driven Ultrasound Video Modeling}

Given an ultrasound video $\mathbf{V}=\{I_t\}_{t=1}^{T}$, we uniformly sample $T$ frames and encode each frame using a pre-trained fetal ultrasound foundation model, i.e., the vision encoder from FetalCLIP~\cite{maani2026fetalclip}. Specifically, the vision encoder is based on a Vision Transformer (ViT) with an input resolution of $224 \times 224$, a patch size of 14, 24 Transformer blocks, and a hidden width of 1024. For each frame $I_t$, the encoder produces a global token $\mathbf{z}^{\mathrm{cls}}_t$, corresponding to the class (CLS) token, together with a sequence of patch tokens $\mathbf{Z}^{\mathrm{patch}}_t=\{\mathbf{z}_{t,n}^{\mathrm{patch}}\}_{n=1}^{N}$, where $\mathbf{z}_{t,n}^{\mathrm{patch}}$ denotes the representation of the $n$-th spatial patch and $N$ is the number of patches in each frame. These two types of representations serve as inputs to different branches of our framework, enabling both global video-level modeling and local spatio-temporal representation learning.

For the main task branch, the global tokens from all sampled frames are arranged as a temporal sequence $\mathbf{Z}^{\mathrm{cls}}=[\mathbf{z}^{\mathrm{cls}}_1,\mathbf{z}^{\mathrm{cls}}_2,\ldots,\mathbf{z}^{\mathrm{cls}}_T] \in \mathbb{R}^{T \times d}$. To encode frame order, we introduce a learnable temporal positional embedding matrix $\mathbf{E}_{\mathrm{temp}} \in \mathbb{R}^{T \times d}$, where each row corresponds to a temporal position. The input to the temporal Transformer is then given by $\widetilde{\mathbf{Z}}^{\mathrm{cls}}=\mathbf{Z}^{\mathrm{cls}}+\mathbf{E}_{\mathrm{temp}}$.
The resulting sequence is then processed by a temporal Transformer encoder $\mathcal{T}(\cdot)$ composed of multi-head self-attention and feed-forward layers to capture inter-frame dependencies.
The temporally encoded frame sequence is defined as $\mathbf{H}^{\mathrm{cls}}=\mathcal{T}(\widetilde{\mathbf{Z}}^{\mathrm{cls}})$. We then apply mean pooling over the temporal dimension to obtain a compact video-level representation, which is formulated as:
\begin{equation}
\mathbf{h}^{\mathrm{vid}}=\frac{1}{T}\sum_{t=1}^{T}\mathbf{H}^{\mathrm{cls}}_t.
\end{equation}

This representation is further transformed by a shared projection block $\phi(\cdot)$ composed of Layer Normalization, a linear layer, and a ReLU activation, yielding $\mathbf{h}=\phi(\mathbf{h}^{\mathrm{vid}})$.
Finally, two task-specific linear heads are built upon $\mathbf{h}$, including a regression head $f_{\mathrm{reg}}(\cdot)$ for ventricular volume estimation and a classification head $f_{\mathrm{cls}}(\cdot)$ for disease prediction:
\begin{equation}
\hat{y}_{\mathrm{reg}} = f_{\mathrm{reg}}(\mathbf{h}), \qquad
\hat{\mathbf{y}}_{\mathrm{cls}} = f_{\mathrm{cls}}(\mathbf{h}).
\end{equation}
Specifically, the regression head is a linear layer mapping $\mathbf{h}\in\mathbb{R}^{d_h}$ to a scalar prediction ($\hat{y}_{\mathrm{reg}}$), while the classification head is a linear layer mapping $\mathbf{h}\in\mathbb{R}^{d_h}$ to a $C$-class logit vector ($\hat{\mathbf{y}}_{\mathrm{cls}}$). 
The regression and classification branches are optimized using a mean squared error (MSE) loss and a class-balanced weighted cross-entropy (CE) loss, controlled by $\lambda_{\mathrm{reg}}$ and $\lambda_{\mathrm{cls}}$:
\begin{equation}
\mathcal{L}_{\mathrm{sup}}=\lambda_{\mathrm{reg}}\mathcal{L}_{\mathrm{MSE}}+\lambda_{\mathrm{cls}}\mathcal{L}_{\mathrm{CE}}.
\end{equation}

\subsection{Tube Latent Prediction for Representation Enhancement}

Although the above supervised regression and classification objectives encourage the model to learn discriminative video-level representations, they impose only limited constraints on the intrinsic local spatio-temporal structure of ultrasound videos. This issue becomes more pronounced when training data and fine-grained annotations are limited, hindering the model from fully exploiting the rich anatomical and dynamic information present in the videos. Inspired by world-model-style representation learning~\cite{ha2018world} and joint-embedding predictive architectures, i.e., JEPA~\cite{assran2023self,bardes2023v,assran2025v}, we encourage the model to infer masked content from visible context in the latent space, thereby learning representations with stronger structural consistency and temporal predictability.

Specifically, we build a tube latent prediction (TLP) objective upon the patch tokens defined above. 
Given the input video $\mathbf{V}$, we first apply spatio-temporal tube masking to obtain a masked video $\widetilde{\mathbf{V}}$. Let $\mathbf{M}\in\{0,1\}^{T\times N}$ denote a binary mask over the patch grid, where $M_{t,n}=1$ indicates that the $n$-th patch token at time step $t$ is masked. To encourage temporal coherence, masking is performed in a tube-wise manner, such that once a spatial patch location is selected, it is masked over a contiguous temporal span. For each frame $I_t$, the masked frame is defined as $\widetilde{I}_t = I_t \odot (1-\Gamma(M_t))$, where $\odot$ denotes element-wise multiplication and $\Gamma(\cdot)$ maps the token-level mask to the corresponding pixel-space masking pattern on the input frame. The masked video is then given by $\widetilde{\mathbf{V}}=\{\widetilde{I}_t\}_{t=1}^{T}$.

Similar to~\cite{munim2026echojepa}, we introduce a teacher-student design for the TLP branch, where the main task branch serves as the \textit{student} encoder ($E_{\mathrm{s}}(\cdot)$).
$\widetilde{\mathbf{V}}$ is then fed into $E_{\mathrm{s}}(\cdot)$, while $\mathbf{V}$ is processed by the \textit{teacher} encoder ($E_{\mathrm{t}}(\cdot)$).
Specifically, the parameters of $E_{\mathrm{t}}(\cdot)$ are updated as an exponential moving average (EMA) of those of $E_{\mathrm{s}}(\cdot)$.
The corresponding student and teacher patch-token representations are given by:
\begin{equation}
\mathbf{Z}^{\mathrm{patch,s}} = E_{\mathrm{s}}(\widetilde{\mathbf{V}}), \qquad
\mathbf{Z}^{\mathrm{patch,t}} = E_{\mathrm{t}}(\mathbf{V}),
\end{equation}

To further capture temporal dependencies across frames for each spatial patch location, we introduce a temporal token encoder $\mathcal{F}_{\mathrm{tok}}(\cdot)$ operating along the temporal dimension. 
The student and teacher patch-token sequences are augmented using a shared learnable temporal positional embedding $\mathbf{E}_{\mathrm{tok}} \in \mathbb{R}^{T \times d_p}$, while the teacher temporal token encoder $\mathcal{F}{\mathrm{tok}}^{\prime}(\cdot)$ is maintained as an EMA copy of the student encoder $\mathcal{F}_{\mathrm{tok}}(\cdot)$.
The temporally contextualized student and teacher token representations are then defined as:
\begin{equation}
\mathbf{H}^{\mathrm{patch,s}} = \mathcal{F}_{\mathrm{tok}}\!\left(\mathbf{Z}^{\mathrm{patch,s}} + \mathbf{E}_{\mathrm{tok}}\right),
\end{equation}
\begin{equation}
\mathbf{H}^{\mathrm{patch,t}} = \mathcal{F}_{\mathrm{tok}}^{\prime}\!\left(\mathbf{Z}^{\mathrm{patch,t}} + \mathbf{E}_{\mathrm{tok}}\right).
\end{equation}

Then, the student token representations are further transformed by a predictor $g(\cdot)$ to obtain the predicted token representations, i.e., $\widehat{\mathbf{H}}^{\mathrm{patch,s}} = g(\mathbf{H}^{\mathrm{patch,s}})$. The TLP objective is computed only over the masked token positions. Let $\Omega=\{(t,n)\mid M_{t,n}=1\}$ denote the set of masked spatio-temporal tokens. The TLP loss is defined as:
\begin{equation}
\mathcal{L}_{\mathrm{TLP}}=\lambda_{\cos}\mathcal{L}_{\cos}+\lambda_{\ell_1}\mathcal{L}_{\ell_1},
\end{equation}
\begin{equation}
\mathcal{L}_{\cos}
=
\frac{1}{|\Omega|}
\sum_{(t,n)\in\Omega}
\left(
1-\cos\left(
\widehat{\mathbf{H}}^{\mathrm{patch,s}}_{t,n},
\mathbf{H}^{\mathrm{patch,t}}_{t,n}
\right)
\right),
\end{equation}
\begin{equation}
\mathcal{L}_{\ell_1}
=
\frac{1}{|\Omega|}
\sum_{(t,n)\in\Omega}
\left\|
\widehat{\mathbf{H}}^{\mathrm{patch,s}}_{t,n}
-
\mathbf{H}^{\mathrm{patch,t}}_{t,n}
\right\|_1.
\end{equation}

\subsection{MRI-Guided Cross-Modal Training Alignment}

During training, we have access to paired ultrasound-MRI data for each case. 
At inference time, only ultrasound videos are available. 
Compared with ultrasound, MRI generally provides clearer visualization of brain anatomy and more stable structural information, although it is substantially more expensive and less accessible in routine clinical practice. 
Therefore, we perform feature-level alignment between the two modalities during training, encouraging the ultrasound encoder to map input videos into an MRI-aligned structural feature space without relying on MRI at test time.

To obtain structurally informative MRI representations, we leverage BOUNTI~\cite{uus2023bounti}, a generalized model for fetal brain MRI. 
Specifically, for each MRI volume, we follow the BOUNTI pipeline and use its pretrained segmentation models, consisting of a U-Net and an Attention U-Net, as strong feature extractors. 
The bottleneck features from the deepest encoder layer of both networks are obtained, globally average-pooled, and concatenated to form a 768-dimensional MRI representation.
To enable cross-modal alignment, we further introduce a lightweight projector $\psi(\cdot)$ to map the MRI representation into the shared embedding space of the ultrasound branch:
\begin{equation}
\mathbf{h}^{\mathrm{MRI}} = \psi(\mathbf{f}^{\mathrm{MRI}}),
\end{equation}
where $\mathbf{f}^{\mathrm{MRI}}$ denotes the original MRI feature and $\mathbf{h}^{\mathrm{MRI}}$ is the projected MRI representation. The projector $\psi(\cdot)$ is implemented as LayerNorm followed by a linear layer and ReLU.
Similarly, we introduce another projector $\varphi(\cdot)$ to transform the ultrasound representation $\mathbf{h}$ into the aligned feature space:
\begin{equation}
\mathbf{h}^{\mathrm{US}} = \varphi(\mathbf{h}),
\end{equation}
where $\mathbf{h}^{\mathrm{US}}$ denotes the aligned ultrasound representation. 
$\varphi(\cdot)$ has the same architecture as $\psi(\cdot)$.
We then perform feature-level alignment between $\mathbf{h}^{\mathrm{US}}$ and $\mathbf{h}^{\mathrm{MRI}}$ in the shared embedding space.
Following common contrastive learning~\cite{radford2021learning}, both features are first normalized, and their pairwise similarities are computed within each mini-batch with size \textit{B}:
\begin{equation}
\mathbf{S}_{ij} = \tau \cdot \mathrm{sim}(\mathbf{h}^{\mathrm{US}}_i,\mathbf{h}^{\mathrm{MRI}}_j),
\end{equation}
where $\mathrm{sim}(\cdot,\cdot)$ denotes cosine similarity and $\tau$ is a learnable scaling factor. Let $\mathbf{P}\in\{0,1\}^{B\times B}$ denote the positive-pair mask, where $P_{ij}=1$ if the $i$-th ultrasound sample and the $j$-th MRI sample belong to the same case, and $P_{ij}=0$ otherwise.
Each row of $\mathbf{P}$ is normalized to form the target distribution:
\begin{equation}
\widetilde{\mathbf{P}}_{ij} = \frac{P_{ij}}{\sum_{k=1}^{B} P_{ik}}.
\end{equation}
The alignment loss is then defined as
\begin{equation}
\mathcal{L}_{\mathrm{align}} = \frac{1}{2}\left(\mathcal{L}_{\mathrm{US}\rightarrow \mathrm{MRI}} + \mathcal{L}_{\mathrm{MRI}\rightarrow \mathrm{US}}\right),
\end{equation}
with
\begin{equation}
\mathcal{L}_{\mathrm{US}\rightarrow \mathrm{MRI}} = - \frac{1}{B}\sum_{i=1}^{B}\sum_{j=1}^{B} \widetilde{\mathbf{P}}_{ij}\log \frac{\exp(\mathbf{S}_{ij})}{\sum_{k=1}^{B}\exp(\mathbf{S}_{ik})},
\end{equation}
and
\begin{equation}
\mathcal{L}_{\mathrm{MRI}\rightarrow \mathrm{US}} = - \frac{1}{B}\sum_{i=1}^{B}\sum_{j=1}^{B} \widetilde{\mathbf{P}}_{ji}\log \frac{\exp(\mathbf{S}_{ji})}{\sum_{k=1}^{B}\exp(\mathbf{S}_{ki})}.
\end{equation}

Through optimizing $\mathcal{L}_{\mathrm{align}}$, the aligned ultrasound representation $\mathbf{h}^{\mathrm{US\rightarrow MRI}}$ is encouraged to capture MRI-consistent structural information. The original ultrasound representation $\mathbf{h}$ is then concatenated with $\mathbf{h}^{\mathrm{US\rightarrow MRI}}$ and further transformed by a lightweight fusion module $\phi_{\mathrm{fuse}}$ (i.e., LayerNorm, Linear, and ReLU) to form the final feature for prediction:
\begin{equation}
\mathbf{h}^{\mathrm{fusion}} = \phi_{\mathrm{fuse}}\left([\mathbf{h};\mathbf{h}^{\mathrm{US\rightarrow MRI}}]\right),
\end{equation}
where $[\cdot;\cdot]$ denotes operation of feature concatenation.

\subsection{VLM Verification and Non-VM Brain Abnormality Alerting}

As mentioned above, video-level training is leveraged to maximize data utilization and facilitate flexible optimization, whereas inference is performed at the case level, where all videos acquired from the same subject are jointly considered to obtain robust predictions. 
Specifically, let a test case contain $K$ ($K\geq1$) videos. For the $k$-th video, the model outputs a regression prediction $\hat{y}^{reg}_k$, a classification logit vector $\hat{\mathbf{y}}^{cls}_k$, and an intermediate feature representation $\mathbf{h}_k \in \mathbb{R}^{d}$. 
We average the multi-video features to obtain the case-level representation $\mathbf{h}_{case}$, which is then fed into the regression and classification heads to obtain $\hat{y}^{reg}_{case}$ and $\hat{\mathbf{y}}^{cls}_{case}$.
The final classification label is determined by applying \textit{argmax} to $\hat{\mathbf{y}}^{cls}_{case}$.

Although case-level fusion improves robustness, uncertainty may still remain due to heterogeneous video quality, incomplete anatomical observations, and inconsistent predictions across views, particularly for underrepresented categories.
To further improve reliability, we introduce a retrieval augmentation strategy that uses clinically similar historical cases to provide useful reference patterns for correcting uncertain predictions. 
Using $\mathbf{h}_{case}$ as the query, we retrieve the top-$M$ most similar cases from the training pool in the learned feature space, forming the retrieval set $\mathcal{R}=\{r_1,\ldots,r_M\}$.

We further employ a training-free VLM as a auxiliary verifier to assess prediction reliability and refine uncertain cases. Specifically, given a query case, we uniformly sample $S$ representative frames from the $K$ available ultrasound videos to construct the query visual set $\{f^{q}_{1,1}, f^{q}_{1,S}, \ldots, f^{q}_{K,S}\}$. 
In this study, $S$ is set to 8 to balance performance and inference speed.
The associated textual prior is provided by the primary model prediction, including the ventricular volume estimate $\hat{y}^{reg}_{case}$ and the predicted diagnostic category $\arg\max(\hat{\mathbf{y}}^{cls}_{case})$.
After retrieving the top-$M$ most similar cases from $\mathcal{R}$, we further collect representative frames from each retrieved case to form the reference visual set $\{f^{r}_{m,1}, \ldots, f^{r}_{m,S}\}_{m=1}^{M}$. The corresponding reference text information includes their known ventricular volumes $\{y^{reg}_{m}\}_{m=1}^{M}$ and diagnostic labels $\{c_m\}_{m=1}^{M}$. 
The query information and retrieved reference evidence are then jointly organized into a multimodal prompt for subsequent VLM-based verification.
Specifically, the prompt template mainly contains the following four components:

\begin{itemize}
\item \textbf{Global instruction:} ``You are an assistant specialized in fetal brain ultrasound analysis. Given the query case and several retrieved reference cases, determine whether the current prediction is clinically reasonable based on visual similarity and diagnostic consistency.''

\item \textbf{Reference content:} 
``Here are several retrieved reference cases.
[\textit{Visual}] Representative ultrasound frames from the retrieved reference case 1. 
[\textit{Text}]: lateral ventricular volume=$y^{reg}_{1}$, diagnosis=$c_1$.'', ..., 
[\textit{Visual}] Representative ultrasound frames from the retrieved reference case $m$. 
[\textit{Text}]: lateral ventricular volume=$y^{reg}_{m}$, diagnosis=$c_m$.''

\item \textbf{Query content:} 
``Here is the current query case.
[\textit{Visual}] Sampled query ultrasound frames $\{f^{q}_{1,1}, \ldots, f^{q}_{K,S}\}$.
[\textit{Text}]: predicted lateral ventricular volume=$\hat{y}^{reg}_{case}$, predicted diagnosis category=$c^{\arg\max(\hat{\mathbf{y}}^{cls}_{case})}$.''

\item \textbf{Question:} 
``Based on the provided reference contents, determine whether the prediction for the current query case is clinically reliable according to visual similarity and diagnostic consistency.
Besides, return a confidence score between 0 and 1 indicating how confident you are in the recalibration suggestion.''
\end{itemize}

After prompting, the VLM outputs a verification decision $d\in\{0,1\}$, where $d=0$ indicates that the current prediction is clinically reliable and the original case-level outputs are directly retained. Otherwise, $d=1$ indicates that calibration is required. 
Then, the VLM further evaluates the relevance of each retrieved reference case $r_m\in\mathcal{R}$ with respect to the query case, and assigns a matching confidence score $\alpha_m \in [0,1]$. The normalized reference weights are computed as
$w_m=\frac{\alpha_m}{\sum_{j=1}^{M}\alpha_j}$.
We then aggregate the retrieved labels using these weights to obtain the reference prediction using the following equations:
\begin{equation}
\mathbf{p}_{ret}=\sum_{m=1}^{M} w_m \mathbf{p}_m,
\end{equation}
\begin{equation}
\hat{y}^{reg}_{ret}=\sum_{m=1}^{M} w_m y^{reg}_m,
\end{equation}
where $\mathbf{p}_m$ and $y^{reg}_m$ denote the one-hot class label distribution and lateral ventricular volume of the $m$-th retrieved case, respectively. 
Then, with the pre-set global adjustment score $\beta=0.3$, the calibrated outputs are computed and updated using Eqs.~\ref{equ:p_final}-\ref{equ:y_final}.
Besides, the final diagnosis category of the query case is determined by applying \textit{argmax} to $\mathbf{p}_{final}$.
\begin{equation}
\label{equ:p_final}
\mathbf{p}_{final}=
\begin{cases}
\mathbf{p}_{case}, & d=0,\\
(1-\beta)\mathbf{p}_{case}+\beta \mathbf{p}_{ret}, & d=1,
\end{cases}
\end{equation}

\begin{equation}
\label{equ:y_final}
\hat{y}^{reg}_{final}=
\begin{cases}
\hat{y}^{reg}_{case}, & d=0,\\
(1-\beta)\hat{y}^{reg}_{case}+\beta \hat{y}^{reg}_{ret}, & d=1.
\end{cases}
\end{equation}

In clinical practice, VM is frequently associated with other fetal brain abnormalities, such as posterior fossa abnormalities (e.g., Dandy-Walker malformation and Blake's pouch cyst), neural tube defects (e.g., Chiari II malformation), and other ventricular system abnormalities beyond the lateral ventricles (e.g., aqueductal stenosis and third/fourth ventricle abnormalities).
However, these abnormalities are challenging to diagnose with common deep learning models, mainly for two reasons. 
First, they often require assessment of diverse anatomical regions across the fetal brain, whereas VM prediction mainly focuses on the lateral ventricles. 
Second, they commonly co-occur with VM or with each other, rather than appearing as isolated findings, complicating the intelligent diagnosis.

Therefore, beyond refining VM-related predictions, it is clinically valuable to further provide alerts for potential non-VM brain abnormalities. 
Motivated by this observation, we explore and extend the retrieval-augmented VLM framework to perform auxiliary abnormality-aware screening.
Specifically, the prompt structure remains unchanged, while the reference content and verification question are further enriched with additional diagnostic descriptions and updated to:

\begin{itemize}
\item \textbf{Reference content$^{*}$:} ``[...] Associated findings=posterior fossa abnormalities (\textit{It refers to structural abnormalities involving the posterior fossa region of the fetal brain. Among them, Dandy-Walker malformation is typically characterized by hypoplasia or agenesis of the cerebellar vermis and enlargement of the posterior fossa.}).''

\item \textbf{Question$^{*}$:} ``[...] In addition, assess whether the query case shows evidence of possible non-VM brain abnormalities. If yes, provide an abnormality alert score between 0 and 1, together with the most likely suspected findings according to the retrieved references.''

\end{itemize}

With the above prompt enhancement, the VLM is able to leverage retrieved abnormal references and the corresponding descriptions to perform auxiliary non-VM abnormality screening in a training-free manner. 
Specifically, it outputs a ranked list of suspected abnormalities, together with abnormality alert scores $\{s_i\}_{i=1}^{N}$ for multiple candidate findings.
Here, $s_i \in [0,1]$ denotes the confidence score of the $i$-th abnormality type. 
All findings satisfying $s_i$$>$$\tau$, where $\tau$ is a predefined threshold, are reported as potential abnormality warnings.

%% file: exp.tex
\begin{figure*}[!t]
    \centering
    \includegraphics[width=1.0\textwidth]{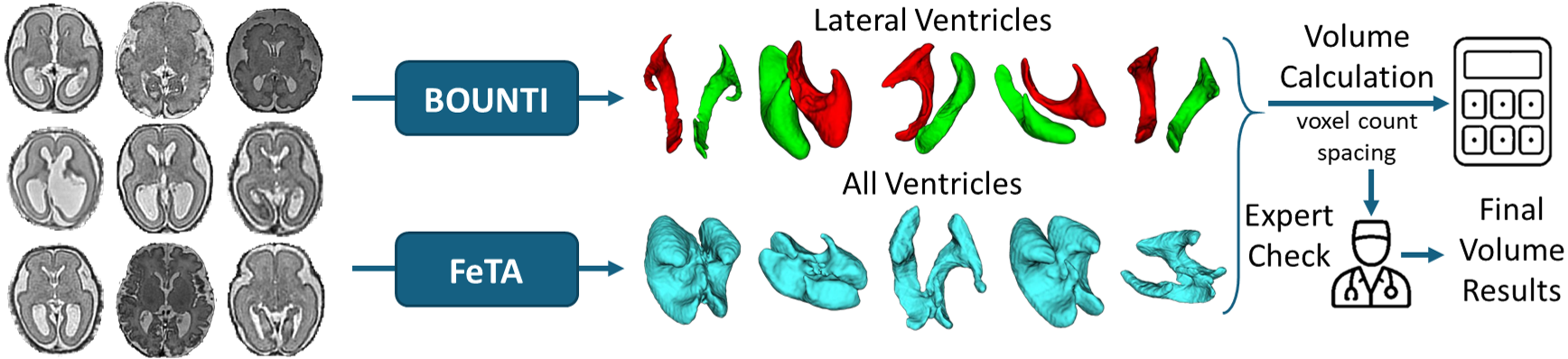}
    \caption{Volume calculation overview in our data preprocessing stage.}
    \label{fig:MRI}
\end{figure*}

\begin{figure*}[!t]
    \centering
    \includegraphics[width=1.0\textwidth]{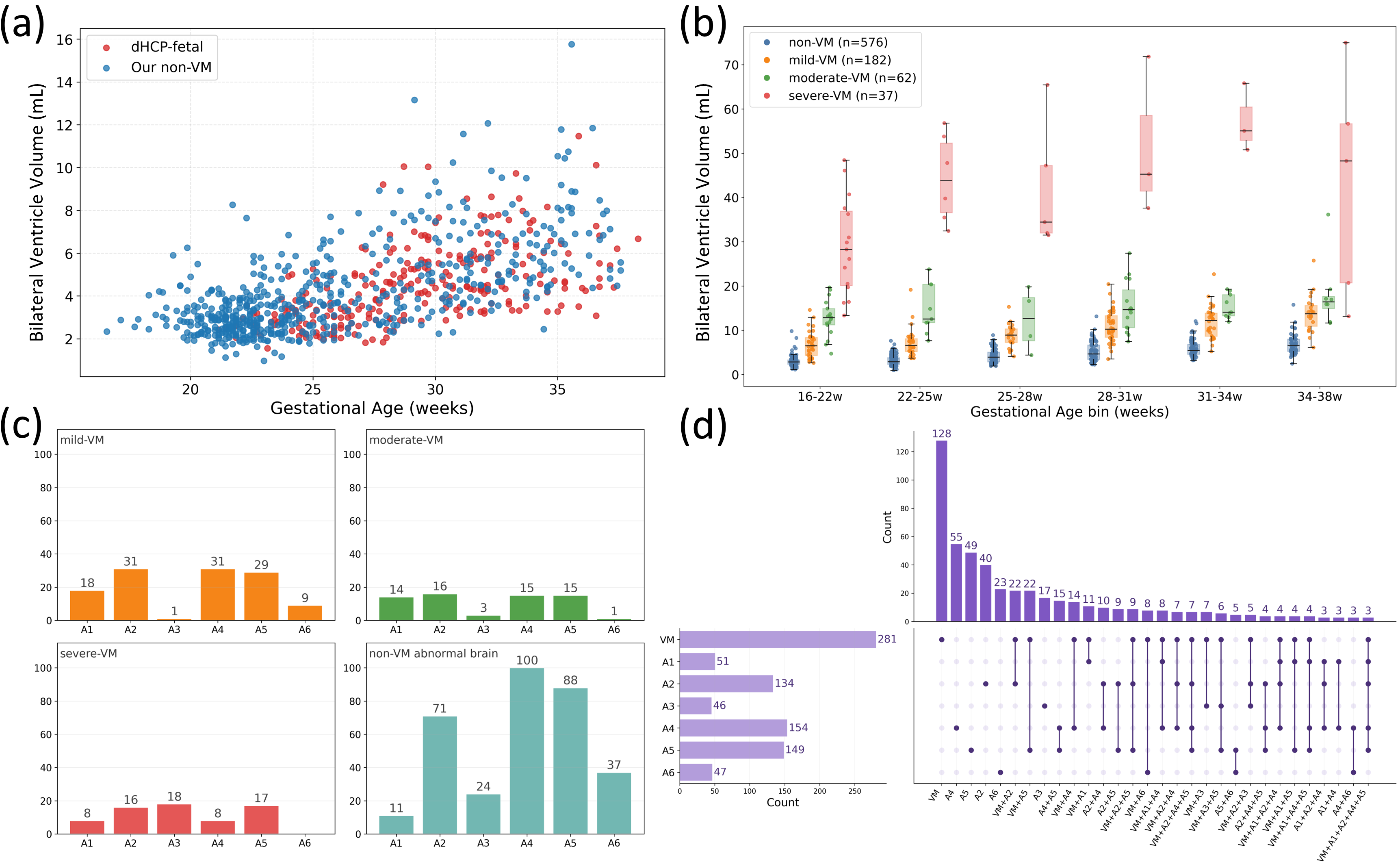}
    \caption{Details of our dataset. 
    (a) bilateral ventricle volume between fetal subjects in the Developing Human Connection Project~\cite{edwards2022developing} (segmentation results generated using DrawEM~\cite{makropoulos2014automatic,makropoulos2018developing} and quality-controlled by the dHCP team) and ours (segmented by deep models~\cite{uus2023bounti,zalevskyi2026advances}).
    Together, the scatter plot analysis and expert visual inspection provide further validation of the accuracy of the automatic segmentation method.
    (b) Statistical distribution of bilateral ventricle volume in non-VM and mild, moderate, and severe VMs across different gestational age groups (weeks).
    (c) Distribution of case counts across six abnormality categories (A1–A6, details refer to Section~\ref{sec:datasets}) among different groups, including non-VM abnormal brain and mild, moderate, and severe VMs.
    (d) The UpSet plot illustrates the set size of each category (VM and A1–A6) and the quantitative distribution of their complex intersections.}
    \label{fig:dataset}
\end{figure*}

\subsection{Datasets}
\label{sec:datasets}
In this study, approved by the Institutional Review Board of Boston Children’s Hospital, we collected a large ultrasound-MRI paired fetal brain dataset for method validation between January 2022 and February 2026.
The dataset comprises 857 cases, each with several stacks of MRI slices for subsequent 3D volume reconstruction and a variable number of ultrasound videos. 
In total, 3,196 videos are included (3.73 videos per case on average). The gestational age (GA) ranges from 16.57 to 38.00 weeks, with a mean of 26.31 and a median of 25.00 weeks. A total of
707, 50, and 100 cases were randomly selected for training, validation, and testing.

In the data preparation pipeline, we first used the NesVoR~\cite{xu2023nesvor} to create the fetal brain MRI volumes, which are subsequently processed using state-of-the-art methods, including BOUNTI~\cite{uus2023bounti} and \href{https://hub.docker.com/r/fetachallenge2024/fetachallenge2024dockerimages}{FeTA}~\cite{zalevskyi2026advances}, to obtain segmentation results of the bilateral lateral ventricles (as shown in Figure~\ref{fig:MRI}).
After careful expert review, the final ground truth (GT) for ventricular volume is computed by summing the segmented voxels and multiplying by the corresponding voxel spacing.
In addition, diagnostic labels (e.g., normal vs. abnormal fetal brain) are extracted from de-identified radiology reports using a large language model (LLM), Qwen3.5-Plus with strong deep reasoning ability~\cite{yang2025qwen3}.
Specifically, the labels include \textit{normal brain (299), mild VM (182), moderate VM (62), severe VM (37), non-VM abnormal brain (277)}.

Based on the radiology reports and LLM analysis, we further provide diagnostic results for various brain abnormalities apart from VM, covering 
A1: neural tube defects (e.g., Spina bifida, Encephalocele), 
A2: midline structure malformations (e.g., Agenesis of the corpus callosum, Holoprosencephaly), 
A3: ventricular system disorders (e.g., Aqueductal stenosis, Hydrocephalus), 
A4: posterior fossa malformations (e.g., Dandy-Walker malformation), 
A5: neurodevelopmental abnormalities (e.g., Microcephaly), and
A6: brain parenchymal abnormalities (e.g., Intracranial cyst, Arachnoid cyst).

Among the 281 VM cases, 128 (45.6\%) were isolated, whereas 84 (29.9\%), 46 (16.4\%), 18 (6.4\%), and 5 (1.8\%) were associated with 1, 2, 3, and 4 additional brain abnormality categories, respectively.
Correspondingly, the numbers of VM cases associated with categories A1-A6 were 40, 63, 22, 54, 61, and 10, respectively.
Among the 277 non-VM abnormal cases, 185, 55, and 12 were associated with 1, 2, and 3 brain abnormality categories, respectively; correspondingly, the numbers of cases assigned to categories A1-A6 were 11, 71, 24, 100, 88, and 37, respectively.
Notably, 25 of the 277 cases belonged to a relatively uncommon group of other brain abnormalities, such as dural sinus malformation, intracranial venous thrombosis, venous sinus malformation, and Vein of Galen malformation, that fell outside the scope of categories A1-A6 and therefore were not considered in our study.
More details about the dataset in this study are provided Figure~\ref{fig:dataset}.
All generated labels are reviewed and verified by a fetal ultrasound expert with over 10 years of clinical experience.

\subsection{Implementation Details}

We implemented VIFBA using Python (v3.9.25) and PyTorch (v2.0.1). All experiments were conducted on a single NVIDIA RTX 6000 Ada Generation GPU with 48 GB memory.
We trained the model for 100 epochs with a batch size of 8 using the AdamW optimizer (learning rate=$1e^{-4}$).
For the MRI projector, we used a reduced learning rate with a scaling factor of 0.1 relative to the main optimizer. 
For ultrasound video processing, 64 frames were uniformly sampled from each video, cropped to remove sensitive information, and resized to $224\times224$. 
The temporal Transformer in the main ultrasound branch consisted of 2 layers with 8 attention heads, an MLP ratio of 4.0, and a dropout rate of 0.1. 
The temporal token encoder in the TLP branch used the same configuration. 
We applied LoRA~\cite{hu2022lora} with rank 8 and $\alpha$=16 to effectively fine-tune the FetalCLIP encoder. 
Gradient checkpointing was enabled during training to reduce memory usage. 
For TLP training, we set the mask ratio to 0.4, the temporal tube length to 4, and the EMA decay for the teacher encoder to 0.996. 
For different loss functions, the weights were set to $\lambda_{\mathrm{reg}}=1.0$, $\lambda_{\mathrm{cls}}=1.0$, $\lambda_{\mathrm{align}}=0.1$, $\lambda_{\cos}=1.0$, and $\lambda_{\ell_1}=0.1$.

\begin{table}[!h]
\centering
\caption{Regression performance across different groups using video-level output aggregation and case-level feature aggregation.
Red subscripts denote the absolute performance gains of case-level feature aggregation over video-level output aggregation. G1: Healthy brain, G2: Non-VM abnormal brain, G3: Mild VM, G4: Moderate VM, and G5: Severe VM.}
\label{tab:group_regression_performance}
\resizebox{\linewidth}{!}{
\begin{tabular}{llccccc}
\toprule
Level & Group &
MSE$\downarrow$ &
RMSE$\downarrow$ &
MAE$\downarrow$ &
Pearson$\uparrow$ &
Spearman$\uparrow$ \\
\midrule
\multirow{6}{*}{Video}
& G1 & 0.2531 & 0.5031 & 0.4390 & 0.9535 & 0.9295 \\
& G2 & 0.7298 & 0.8543 & 0.7338 & 0.9618 & 0.8809 \\
& G3 & 2.5167 & 1.5864 & 1.4416 & 0.9208 & 0.8684 \\
& G4 & 7.0826 & 2.6613 & 2.2956 & 0.9611 & 1.0000 \\
& G5 & 7.6917 & 2.7734 & 2.5630 & 0.9479 & 1.0000 \\
& Overall & 1.6476 & 1.2836 & 0.9364 & 0.9793 & 0.9403 \\
\midrule
\multirow{6}{*}{Case}
& G1
& 0.0473\gain{0.2058}
& 0.2175\gain{0.2856}
& 0.1985\gain{0.2405}
& 0.9916\gain{0.0381}
& 0.9685\gain{0.0390} \\

& G2
& 0.2327\gain{0.4971}
& 0.4824\gain{0.3719}
& 0.4341\gain{0.2997}
& 0.9879\gain{0.0261}
& 0.9217\gain{0.0408} \\

& G3
& 1.3568\gain{1.1599}
& 1.1648\gain{0.4216}
& 1.0389\gain{0.4027}
& 0.9526\gain{0.0318}
& 0.9351\gain{0.0667} \\

& G4
& 2.8950\gain{4.1876}
& 1.7015\gain{0.9598}
& 1.5207\gain{0.7749}
& 0.9842\gain{0.0231}
& 1.0000\same{0.0000} \\

& G5
& 4.2625\gain{3.4292}
& 2.0646\gain{0.7088}
& 1.8706\gain{0.6924}
& 0.9707\gain{0.0228}
& 1.0000\same{0.0000} \\

& Overall
& \textbf{0.7507}\gain{0.8969}
& \textbf{0.8664}\gain{0.4172}
& \textbf{0.5909}\gain{0.3455}
& \textbf{0.9907}\gain{0.0114}
& \textbf{0.9730}\gain{0.0327} \\
\bottomrule
\end{tabular}}
\end{table}

The proposed VIFBA model was evaluated on both regression and classification tasks.
For volume regression, performance was assessed using mean absolute error (MAE), root mean square error (RMSE), Mean Squared Error (MSE), Pearson’s correlation coefficient, and Spearman’s rank correlation coefficient.
For VM evaluation, performance was evaluated using accuracy (Acc), macro-precision (Pre), macro-recall (Rec), macro-F1 (F1), and macro-AUC (AUC).
For multi-label classification, performance was evaluated using Pre, Rec, F1, Hamming loss (HL), and exact match ratio (EMR).
Specifically, HL and EMR are formally defined as:

\begin{equation}
\mathrm{HL}=\frac{1}{N C}
\sum_{i=1}^{N}\sum_{j=1}^{C}
\mathbb{I}(y_{ij}\neq \hat{y}_{ij}),
\end{equation}

\begin{equation}
\mathrm{EMR}=\frac{1}{N}
\sum_{i=1}^{N}
\mathbb{I}
\left(
\mathbf{y}_{i}=\hat{\mathbf{y}}_{i}
\right),
\end{equation}
where $N$ denotes the number of samples and $C$ denotes the number of labels. 
$\mathbf{y}_{i}$ and $\hat{\mathbf{y}}_{i}$ represent the ground-truth and predicted label vectors of the $i$-th sample, respectively. 
$\mathbb{I}(\cdot)$ is the indicator function, which returns 1 when the condition is satisfied and 0 otherwise.
Lower HL and higher EMR indicate better multi-label classification performance.

Unless otherwise specified, all experiments were conducted using fixed random seeds for fair comparison. 
In addition, to better assess the robustness of different methods across random initializations and to evaluate the statistical significance of performance differences, selected experiments were repeated with 10 different random seeds.
For statistical analysis, we first computed the paired differences between two methods across the 10 repeated runs and assessed their normality using the Shapiro-Wilk test.
If the paired differences satisfied the normality assumption, a one-sided paired $t$-test was used to examine whether the target method significantly outperformed the compared method.
Otherwise, the Wilcoxon signed-rank test was used as a non-parametric alternative for the same one-sided paired comparison.
For MSE, RMSE, and MAE, significance was tested in the direction of lower values, whereas for Pearson, Spearman, and classification metrics, significance was tested in the direction of higher values.
Statistical significance was determined at $p<0.05$.

\begin{table}[!t]
  \centering
  \caption{Method Comparison on the regression task.}
    \begin{tabular}{cccccc}
    \toprule
          & MSE   & RMSE  & MAE   & Pearson & Spearman  \\
    \midrule
    I3D   & 12.6789  & 3.5607  & 2.7929  & 0.8470  & 0.6547  \\
    R(2+1)D & 13.0733  & 3.6157  & 2.5684  & 0.8243  & 0.6678  \\
    TimeSformer & 11.8981  & 3.4494  & 2.6758  & 0.8906  & 0.7331  \\
    InternVideo2 & 11.4605  & 3.3853  & 2.5386  & 0.9510  & 0.8318  \\
    VideoMamba & 12.0456  & 3.4707  & 2.6615  & 0.8752  & 0.7496  \\
    V-JEPA2 & 10.8888  & 3.2998  & 2.6546  & 0.8821  & 0.6937  \\
    \midrule
    Baseline (Reg) & 9.9199  & 3.1496  & 2.6122  & 0.9033  & 0.7648  \\
    VIFBA w/o Cls & 3.2208  & 1.7947  & 1.4461  & 0.9849  & 0.9562  \\
    VIFBA  & \textbf{0.7507}  & \textbf{0.8664}  & \textbf{0.5909}  & \textbf{0.9907}  & \textbf{0.9730}  \\
    \bottomrule
    \end{tabular}%
  \label{tab:result-reg}%
\end{table}%

\begin{table}[!t]
  \centering
  \caption{Method Comparison on the classification task.}
    \begin{tabular}{cccccc}         
    \toprule & Acc   & Pre   & Rec   & F1    & AUC \\
    \midrule
    I3D   & 0.7000  & 0.6361  & 0.7611  & 0.6610  & 0.8899  \\
    R(2+1)D & 0.7000  & 0.6264  & 0.7086  & 0.6450  & 0.8659  \\
    TimeSformer & 0.7300  & 0.6462  & 0.7196  & 0.6666  & 0.9006  \\
    InternVideo2 & 0.7500  & 0.6605  & 0.7545  & 0.6762  & 0.9030  \\
    VideoMamba & 0.7300  & 0.6498  & 0.7424  & 0.6670  & 0.8996  \\
    V-JEPA2 & 0.7700  & 0.7247  & 0.7604  & 0.7208  & 0.9199  \\
    \midrule
    Baseline (Cls) & 0.7900  & 0.7387  & 0.8256  & 0.7568  & 0.9186  \\
    VIFBA w/o Reg & 0.8600  & 0.8075  & 0.8574  & 0.8173  & 0.9500  \\
    VIFBA  & \textbf{0.9400}  & \textbf{0.9357}  & \textbf{0.9068}  & \textbf{0.9130}  & \textbf{0.9793}  \\
    \bottomrule
    \end{tabular}%
  \label{tab:result-cls}%
\end{table}%

\begin{figure*}[!t]
\centering
\includegraphics[width=1.0\textwidth]{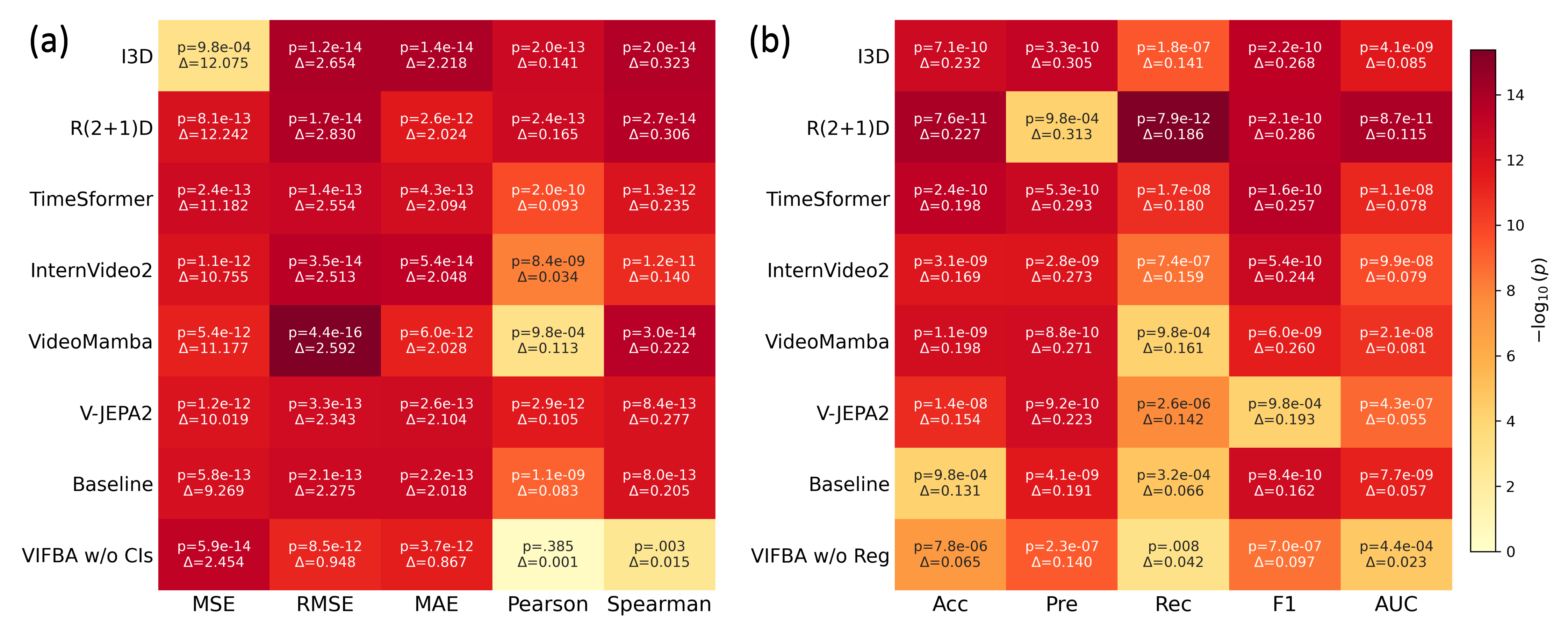}
\caption{
Comparisons with statistical significance analysis between VIFBA and competitors on (a) regression and (b) classification tasks. Each cell reports the $p$-value and the mean gain $\Delta$ of VIFBA.
}  
\label{fig:p-value1}
\end{figure*}

\begin{figure*}[!t]
    \centering
    \includegraphics[width=1.0\textwidth]{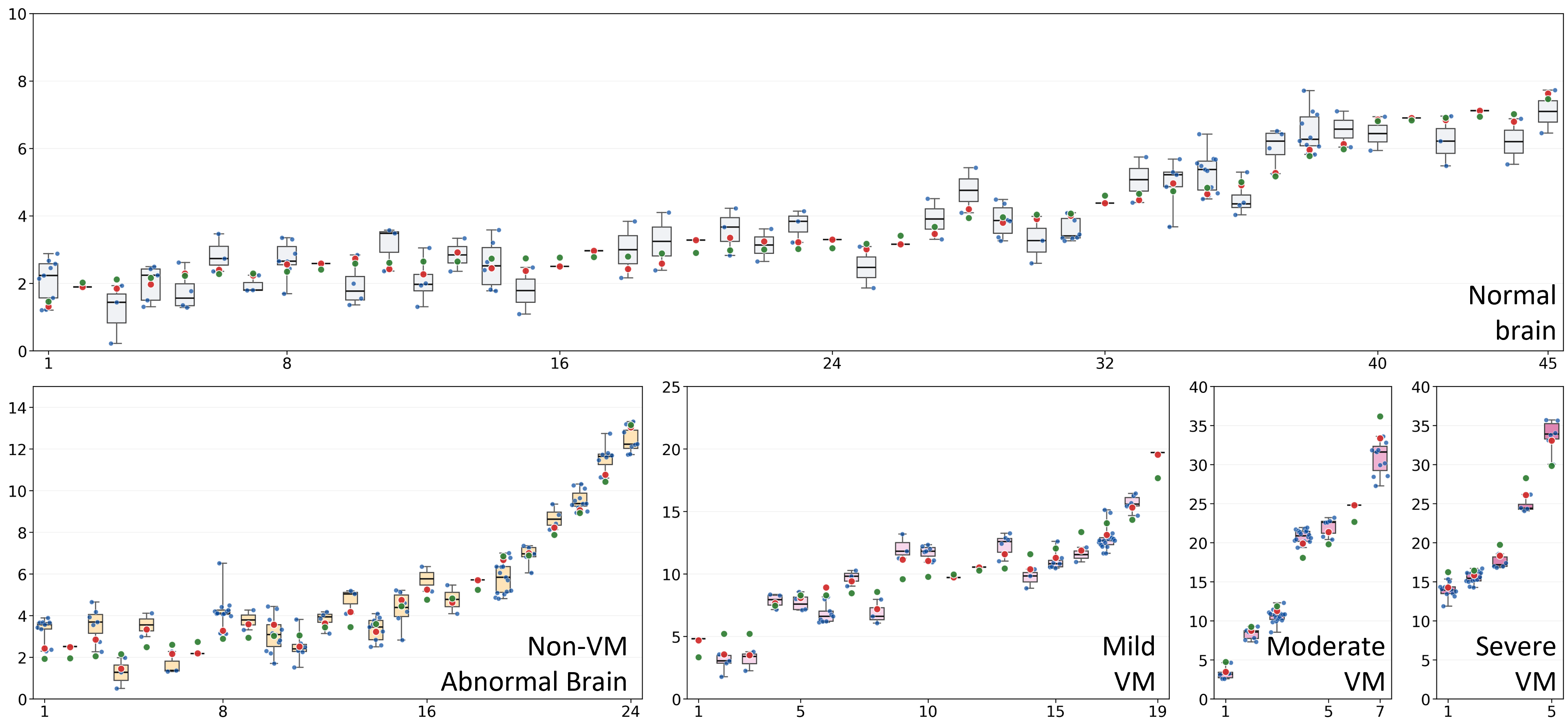}
    \caption{Group-wise detailed regression performance, including normal brain (n=45), non-VM abnormal brain (n=24), mild VM (n=19), moderate VM (n=7), and severe VM (n=5).
    Y-axis represents lateral ventricular volume (mL).
    For each group, cases are sorted by the ground-truth lateral ventricular volume (\textcolor{green}{green circles}).
    The \textcolor{blue}{blue circles} indicate the video-level predictions, while the \textcolor{red}{red circles} show the case-level output.}  
    \label{fig:results_group}
\end{figure*}

\subsection{Comparison with Video-Based Methods}

Tables~\ref{tab:result-reg} and~\ref{tab:result-cls} compare the proposed method with representative video-based models on lateral ventricular volume regression and VM severity classification, respectively.
Specifically, we selected six representative and state-of-the-art video analysis methods as competitors (I3D~\cite{carreira2017quo}, R(2+1)D~\cite{tran2018closer}, TimeSformer~\cite{bertasius2021space}, InternVideo2~\cite{wang2024internvideo2}, VideoMamba~\cite{li2024videomamba}, and V-JEPA2~\cite{assran2025v}), and replaced their original heads with a regression head or classification head for ventricular volume estimation and VM severity prediction, respectively.

\begin{table*}[!t]
  \centering
  \caption{Ablation study on our VIFBA framework.}
    \begin{tabular}{ccccccccccccc}
    \toprule
    \multicolumn{3}{c}{\multirow{2}[4]{*}{Method \& Module}} & \multicolumn{5}{c}{Regression}        & \multicolumn{5}{c}{Classification} \\
\cmidrule{4-13}    \multicolumn{3}{c}{}  & MSE   & RMSE  & MAE   & Pearson & Spearman  & Acc   & Pre   & Rec   & F1    & AUC \\
    \midrule
    \multicolumn{3}{c}{Baseline (Reg)} & 9.9199 & 3.1496 & 2.6122 & 0.9033 & 0.7648 & /     & /     & /     & /     & / \\
    \multicolumn{3}{c}{Baseline (Cls)} & /     & /     & /     & /     & /     & 0.7900  & 0.7387  & 0.8256  & 0.7568  & 0.9186  \\

    \multicolumn{3}{c}{Baseline (Reg+Cls)} & 6.3064 & 2.5112 & 2.0827 & 0.9707 & 0.9304 & 0.8200  & 0.7310  & 0.8423  & 0.7569  & 0.9495  \\
    \midrule
    TLP  & CMTA & VLM   & MSE   & RMSE  & MAE   & Pearson & Spearman  & Acc   & Pre   & Rec   & F1    & AUC \\
    \midrule
    \ding{51}     &   \ding{55}    &   \ding{55}    & 2.1778  & 1.4757  & 1.0739  & 0.9882  & 0.9179  & 0.8800  & 0.8002  & 0.8890  & 0.8302  & 0.9601  \\
       \ding{55}   & \ding{51}     &  \ding{55}     & 2.3407  & 1.5299  & 1.2608  & 0.9759  & 0.9048  & 0.8700  & 0.7799  & 0.8451  & 0.8037  & 0.9498  \\
      \ding{55}   &   \ding{55}    & \ding{51}     & 3.8378  & 1.9590  & 1.5638  & 0.9652  & 0.8643  & 0.8200  & 0.7344  & 0.8211  & 0.7639  & 0.9513  \\
    \ding{51}     & \ding{51}     &  \ding{55}     & 1.3936  & 1.1805  & 0.9140  & 0.9844  & 0.9094  & 0.8900  & 0.8717  & 0.9070  & 0.8863  & 0.9602  \\
    \ding{51}     &    \ding{55}   & \ding{51}     & 1.7897  & 1.3378  & 1.0914  & 0.9837  & 0.9630  & 0.9100  & 0.9144  & 0.8706  & 0.8814  & 0.9648  \\
      \ding{55}    & \ding{51}     & \ding{51}     & 1.2619  & 1.1223  & 0.9260  & 0.9875  & 0.9326  & 0.9000  & 0.8223  & 0.8995  & 0.8474  & 0.9612  \\
    \midrule
    \multicolumn{3}{c}{VIFBA} & \textbf{0.7507}  & \textbf{0.8664}  & \textbf{0.5909}  & \textbf{0.9907}  & \textbf{0.9730}  & \textbf{0.9400}  & \textbf{0.9357}  & \textbf{0.9068}  & \textbf{0.9130}  & \textbf{0.9793}  \\
    \bottomrule
    \end{tabular}%
  \label{tab:result-ours}%
\end{table*}%

\begin{table*}[!t]
  \centering
  \caption{Comparison of VIFBA with different foundation model-based encoders.}
    \begin{tabular}{ccccccccccc}
    \toprule
    \multicolumn{1}{c}{\multirow{2}[4]{*}{VIFBA}} & \multicolumn{5}{c}{Regression}        & \multicolumn{5}{c}{Classification} \\
\cmidrule{2-11}          & MSE   & RMSE  & MAE   & Pearson & Spearman  & Acc   & Pre   & Rec   & F1    & AUC \\
    \midrule
    PMC-CLIP & 1.1804  & 1.0864  & 0.8970  & 0.9893  & 0.9239  & 0.8900  & 0.8398  & 0.8917  & 0.8537  & 0.9296  \\
    BiomedCLIP & 1.0475  & 1.0234  & 0.8300  & 0.9915  & 0.9466  & 0.9100  & 0.8652  & 0.9122  & 0.8849  & 0.9572  \\
    USFM  & 0.8702  & 0.9329  & 0.7904  & \textbf{0.9965}  & 0.9704  & \textbf{0.9400}  & 0.8878  & 0.9300  & 0.9053  & 0.9717  \\
    EchoCare & 0.9217  & 0.9600  & 0.8049  & 0.9905  & 0.9400  & 0.9300  & 0.8793  & 0.8523  & 0.8580  & 0.9399  \\
    Ultrasound-CLIP & 0.9013  & 0.9494  & 0.7994  & 0.9942  & 0.9376  & 0.9300  & 0.8747  & \textbf{0.9550}  & 0.9048  & 0.9717  \\
    FetalCLIP & \textbf{0.7507}  & \textbf{0.8664}  & \textbf{0.5909}  & 0.9907  & \textbf{0.9730}  & \textbf{0.9400}  & \textbf{0.9357}  & 0.9068  & \textbf{0.9130}  & \textbf{0.9793}  \\
    \bottomrule
    \end{tabular}%
  \label{tab:result-encoder}%
\end{table*}%

For volume regression, all compared methods showed limited performance, with MSE values ranging from 10.8888 to 13.0733.
\textit{Baseline (Reg)} with FetalCLIP backbone reduced the MSE to 9.9199 and achieved a Pearson correlation of 0.9033. 
Compared with them, VIFBA achieved substantially better performance, reducing the MSE, RMSE, and MAE to 0.7507, 0.8664, and 0.5909, respectively. 
It also achieved Pearson and Spearman correlation coefficients of 0.9907 and 0.9730, respectively, indicating strong agreement with the MRI-derived ventricular volume measurements.
We fairly removed the classification branch and found that \textit{VIFBA w/o Cls} outperforms all compared models and the regression baseline.

For the VM severity classification task, the selected video models achieved Accs between 0.7000 and 0.7700, with F1 ranging from 0.6450 to 0.7208.
Specifically, V-JEPA2 obtained the best classification performance among them, with an Acc of 0.7700 and an F1 of 0.7208. 
\textit{Baseline (Cls)} achieved higher performance, with an Acc of 0.7900 and a F1 score of 0.7568. 
In contrast, VIFBA achieved the best overall classification results, with an Acc of 0.9400, Pre of 0.9357, F1 of 0.9130, and AUC of 0.9793. 
For comprehensive comparison, VIFBA w/o Reg also outperformed the classification baseline, achieving an Acc of 0.8600 and an F1 of 0.8173.

Moreover, statistical analysis based on multi-seed experiments is presented in Figure~\ref{fig:p-value1}, further confirming the significant performance improvements achieved by VIFBA.
These results indicate that general model designs for video analysis are insufficient for fine-grained fetal brain ultrasound predictions. 
In contrast, VIFBA learns more effective video representations through foundation model-based encoding, latent predictive learning, and MRI-informed supervision, while the gains over single-task variants suggest the complementarity between volume estimation and VM classification.

We also provide the video- and case-level regression performance on our test set, as shown in Figure~\ref{fig:results_group}.
For each group, cases were sorted by the ground-truth lateral ventricular volume.
Overall, the model predictions exhibit good agreement with the ground truth across normal brain, non-VM abnormal brain, and different VM severity groups.
It can be seen that video-level predictions may vary due to differences in imaging views, video quality, and incomplete anatomical observations.
By aggregating multi-video features at the case level, the model integrates complementary anatomical information and produces more stable volume estimates that better follow the ground-truth trend across severity levels.
Moreover, as reported in Table~\ref{tab:group_regression_performance}, our case-level feature aggregation strategy consistently outperforms video-level output aggregation across all diagnostic subgroups and the entire test set.

\begin{figure*}[!t]
\centering
\includegraphics[width=1.0\textwidth]{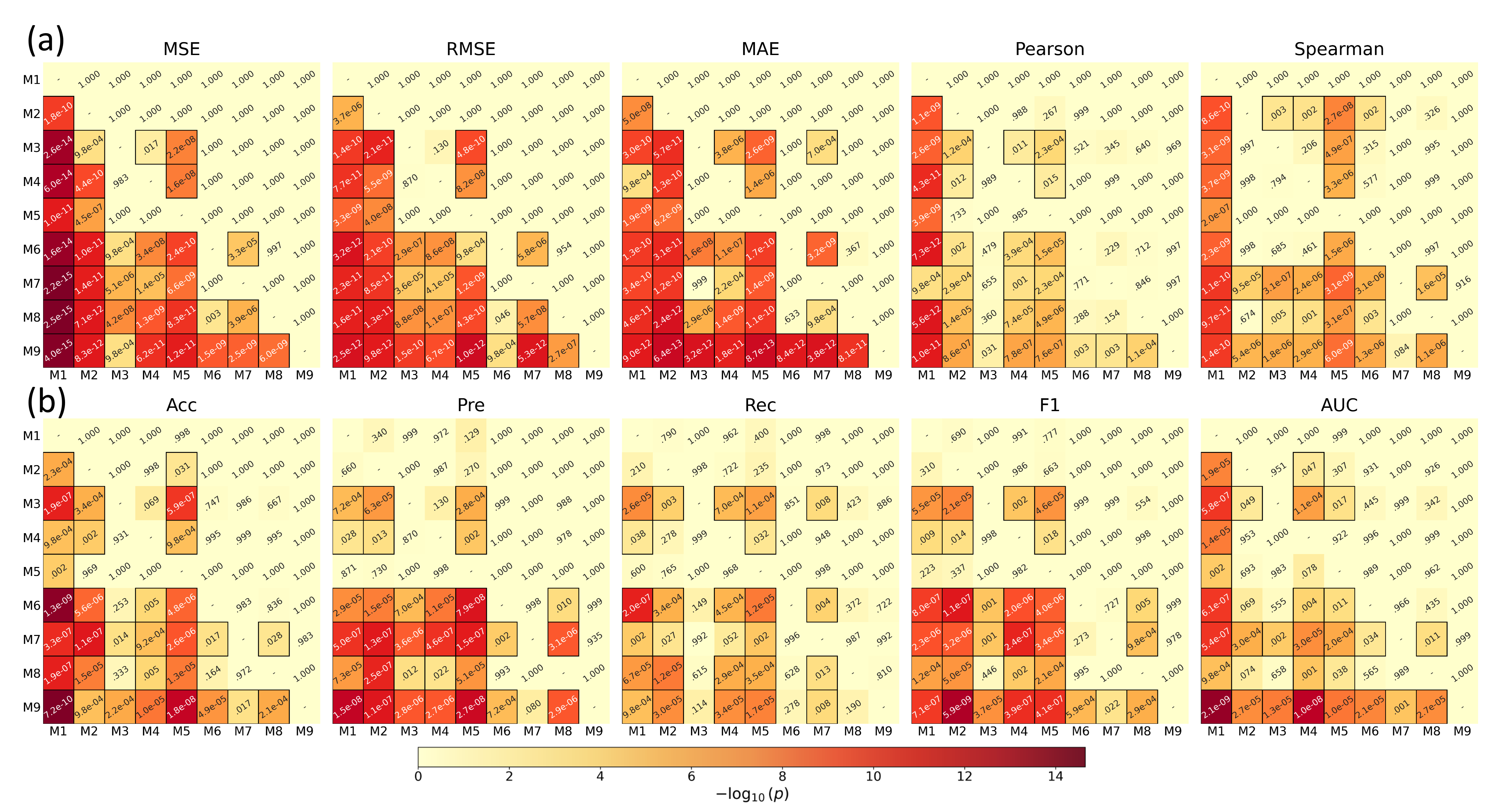}
\caption{
Pairwise statistical results among ablation variants on (a) regression and (b) classification metrics.
Each cell reports the $p$-value for testing whether the row method significantly outperforms the column method.
Boxes with black borders indicate statistically significant improvements ($p<0.05$).
M1: Baseline (Reg)/(Cls), M2: Baseline (Reg+Cls), M3: TLP, M4: CMTA, M5: VLM, M6: TLP+CMTA, M7: TLP+VLM, M8: CMTA+VLM, M9: VIFBA.
}  
\label{fig:p-value2}
\end{figure*}

\subsection{Ablation Study}

We further conducted ablation experiments to investigate the contribution of each component, including tube latent prediction (TLP), cross-modal training alignment (CMTA), and abnormality verification \& alerting (VLM). 
As shown in Table~\ref{tab:result-ours}, each component consistently improved the baseline. When used individually, the TLP objective achieved an MSE of 2.1778, an MAE of 1.0739, and an Acc of 0.8800, suggesting that it benefits video representation learning. 
CMTA also improved both regression and classification performance, achieving an MSE of 2.3407, an MAE of 1.2608, and an Acc of 0.8700, suggesting that MRI-derived representations provide complementary supervision during training. 
The VLM module alone also enhanced model performance, especially for regression (MSE reduction: 2.4686; MAE reduction: 0.5189).

Combining different modules further improved performance. 
For example, the combination of TLP and CMTA reduced the MSE and MAE to 1.3936 and 0.9140, respectively, and improved the F1 to 0.8863. 
Connecting TLP with VLM achieved a higher Acc of 0.9100, while maintaining strong regression metrics with an MSE of 1.7897 and an MAE of 1.0914. 
The full VIFBA achieved the best overall performance, with an MSE of 0.7507, an MAE of 0.5909, a Pearson correlation of 0.9907, an Acc of 0.9400, and an F1 of 0.9130.

We further repeated the experiments using 10 different random seeds and report the corresponding statistical $p$-values in Figure~\ref{fig:p-value2}.
The statistical analysis shows that the proposed modules yield statistically significant improvements in most comparisons, although the gains from individual modules are relatively limited in some settings and do not consistently reach significance across all metrics.
Notably, the numerous significant improvements of VIFBA over different ablation variants, as highlighted in the last row of each matrix, further demonstrate the complementarity of the three components.

\begin{figure*}[!t]
    \centering
    \includegraphics[width=1.0\textwidth]{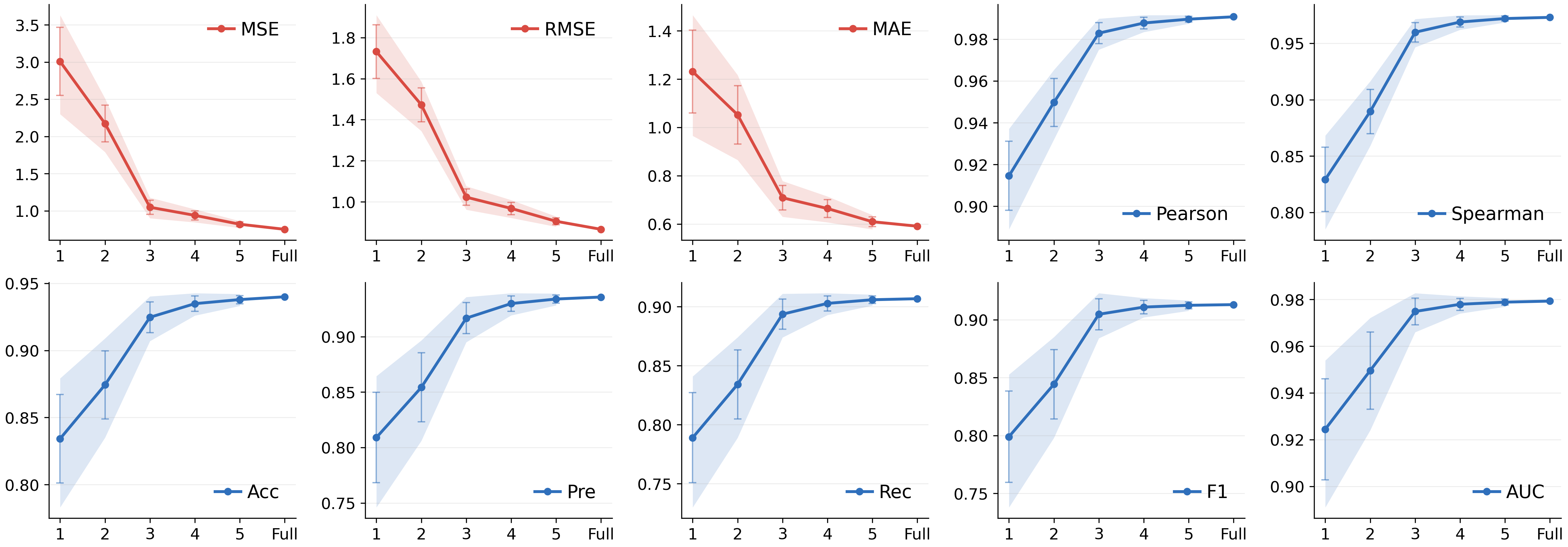}
    \caption{Performance variation with different numbers of selected videos per case. 
    Full denotes using all available videos for case-level feature aggregation.}  
    \label{fig:video-num}
\end{figure*}

\subsection{Impact of Different Numbers of Videos}

To investigate the impact of the number of ultrasound videos used for case-level feature aggregation, we conducted an ablation study by varying the number of selected videos per case, as shown in Figure~\ref{fig:video-num}.
Since each case may contain a different number of brain ultrasound videos, we randomly sampled a predefined number (e.g., 1--5) from each case. 
If the available number was smaller than the specified number, all available videos were used. To reduce the influence of random video selection, the sampling process was repeated five times for each setting. The curves represent the mean performance, the shaded regions indicate the minimum-to-maximum range, and the error bars denote the standard deviation.

It can be seen that increasing the number of selected videos consistently improves the regression and classification performance.
Most improvements are achieved when increasing the number of videos from 1 to 3, after which the performance gradually converges.
Meanwhile, the shaded regions become narrower with more available videos, indicating reduced prediction variability and improved stability of case-level feature aggregation. 
These overall results suggest that aggregating multiple videos enables our VIFBA to capture complementary anatomical information from different scanning views and provides more robust case-level predictions.

\begin{table}[!t]
  \centering
  \caption{Comparison of VIFBA with baselines and different VLMs on the multi-label classification task. S1: one-stage classification over 10 categories (normal brain, mild/moderate/severe VM, and A1–A6); S2: two-stage classification that first identifies abnormal brains and then determines their specific diseases.}
   \resizebox{\linewidth}{!}{
    \begin{tabular}{ccccccc}
    \toprule
    \multicolumn{2}{c}{Method} & Pre   & Rec   & F1    & HL    & EMR \\
    \midrule
    \multicolumn{2}{c}{VIFBA-CLS-S1} & 0.4243  & 0.6299  & 0.4575  & 0.1250  & 0.3200  \\
    \multicolumn{2}{c}{VIFBA-CLS-S2} & 0.4762  & 0.7036  & 0.5084  & 0.1110  & 0.4100  \\
    \midrule
    \multirow{9}[2]{*}{VIFBA} & Claude-Haiku-4.5 & 0.6129  & 0.7836  & 0.6673  & 0.0630  & 0.6900  \\
          & Gemini-3.1-Pro & 0.6517  & 0.8027  & 0.7053  & 0.0530  & 0.7000  \\
          & Qwen3.5-Plus & 0.6485  & 0.7847  & 0.6861  & 0.0590  & 0.7100  \\
          & GPT-5.4 & 0.6333  & 0.8346  & 0.6941  & 0.0610  & 0.7100  \\
          & Qwen3.6-Plus & 0.6866  & 0.8384  & 0.7332  & 0.0530  & 0.7200  \\
          & Qwen3.7-Plus & 0.6955   &  0.8595  &  0.7478 & 0.0500  &  0.7300 \\
          & GPT-5.5 & 0.7090  & 0.8999  & 0.7617  & 0.0450  & 0.7300  \\
          & GPT-5.6 Terra & 0.7130  & 0.8775  & 0.7687  & 0.0450  & 0.7400 \\
          & GPT-5.6 Sol & \textbf{0.7207}  & \textbf{0.8933}  & \textbf{0.7764}  & \textbf{0.0440}  & \textbf{0.7500} \\
    \bottomrule
    \end{tabular}}%
  \label{tab:result-vlm}%
\end{table}%

\begin{figure*}[!t]
    \centering
    \includegraphics[width=1.0\textwidth]{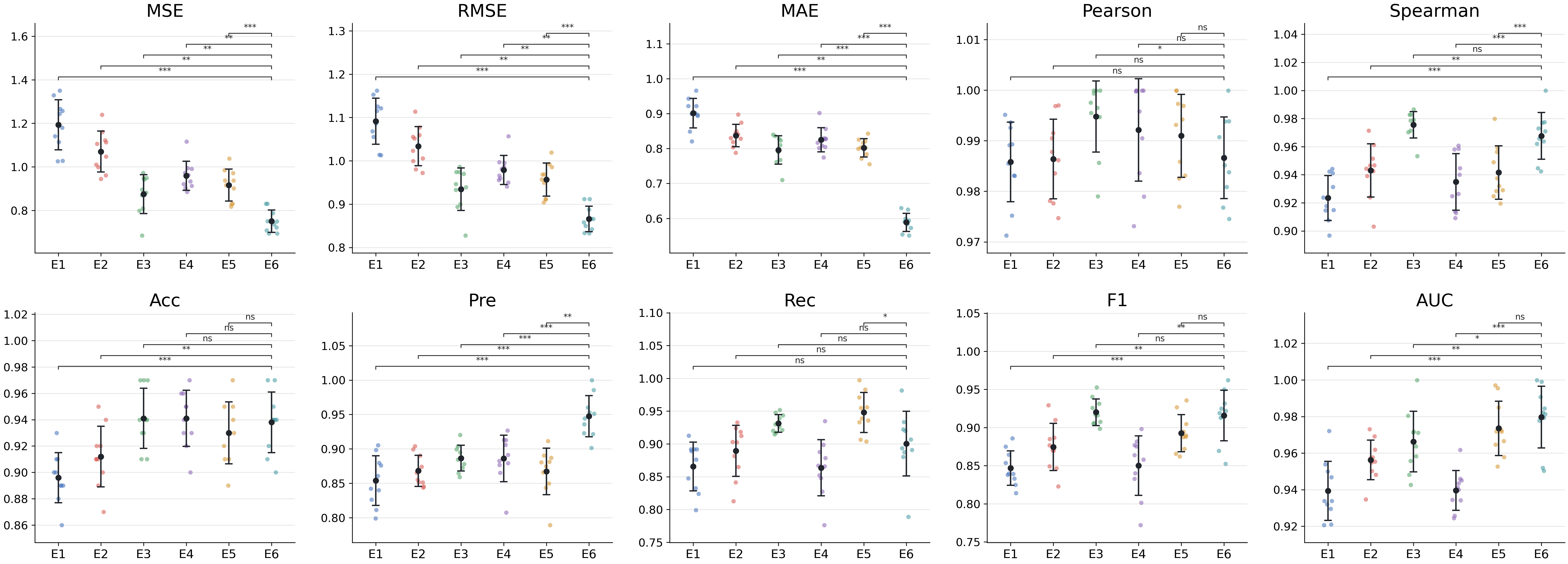}
    \caption{Multi-seed performance comparison of VIFBA with different foundation encoders.
    E1--E6 represent PMC-CLIP, BiomedCLIP, USFM, EchoCare, Ultrasound-CLIP, and FetalCLIP, respectively.
    Statistical significance is denoted as * ($p<0.05$), ** ($p<0.01$), *** ($p<0.001$), and ns (not significant).
    }  
    \label{fig:encoder_10}
\end{figure*}

\begin{figure*}[!t]
    \centering
    \includegraphics[width=1.0\textwidth]{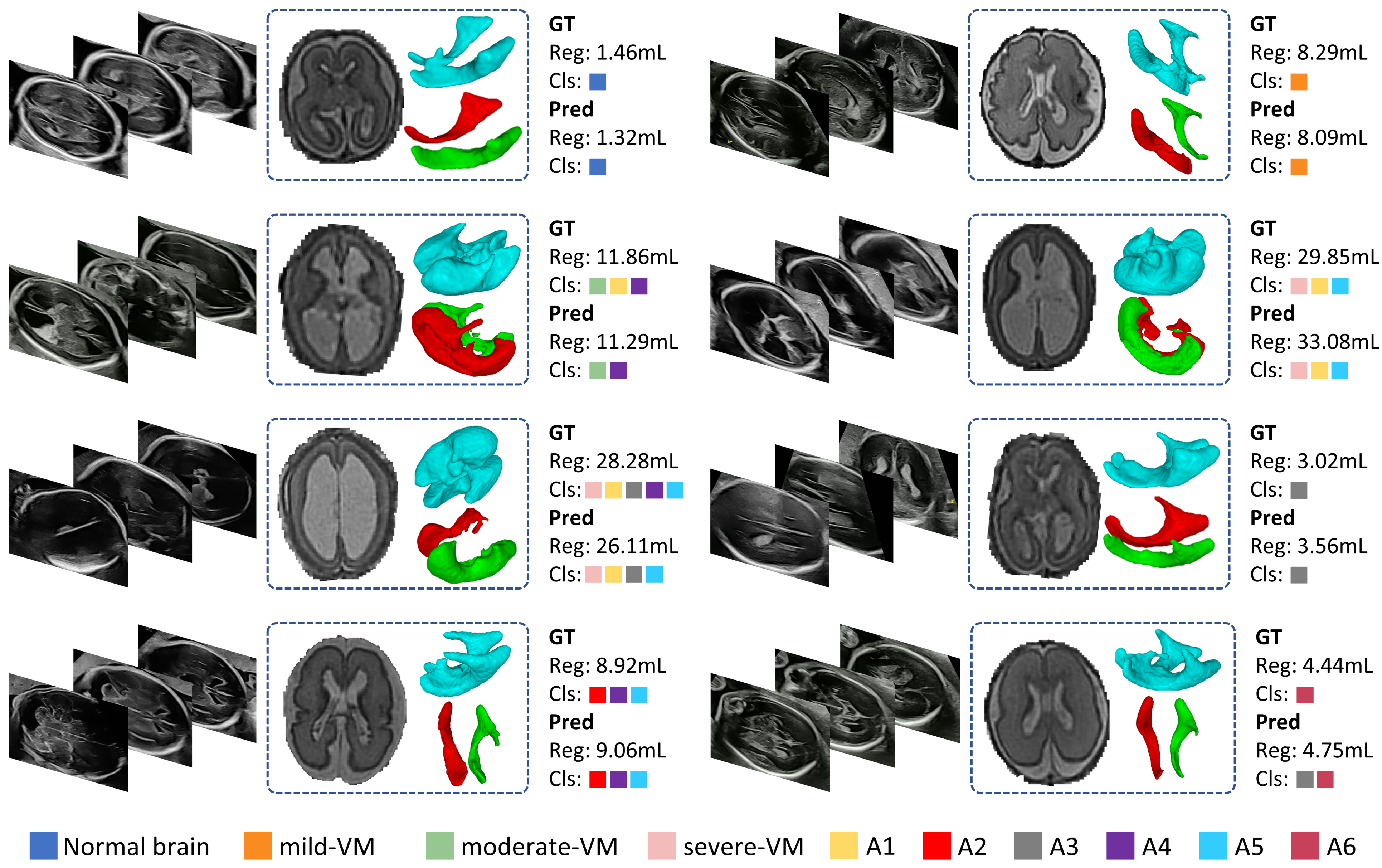}
    \caption{Visualization of selected test cases. For each case, the figure shows example frames from the input ultrasound videos, an axial slice from the reconstructed MRI volume, and ventricular segmentation results obtained from the MRI based on BOUNTI and FeTA segmentors. 
    Note that, to facilitate visualization of the ventricular shape, we have arbitrarily rotated the ventricular segmentation masks for each case. In addition, we show the ground-truth (GT, based on the segmentations/reports and checked by an expert) and VIFBA-predicted (pred) ventricular volumes and diagnostic labels.} 
    \label{fig:case}
\end{figure*}

\begin{figure*}[!t]
    \centering
    \includegraphics[width=1.0\textwidth]{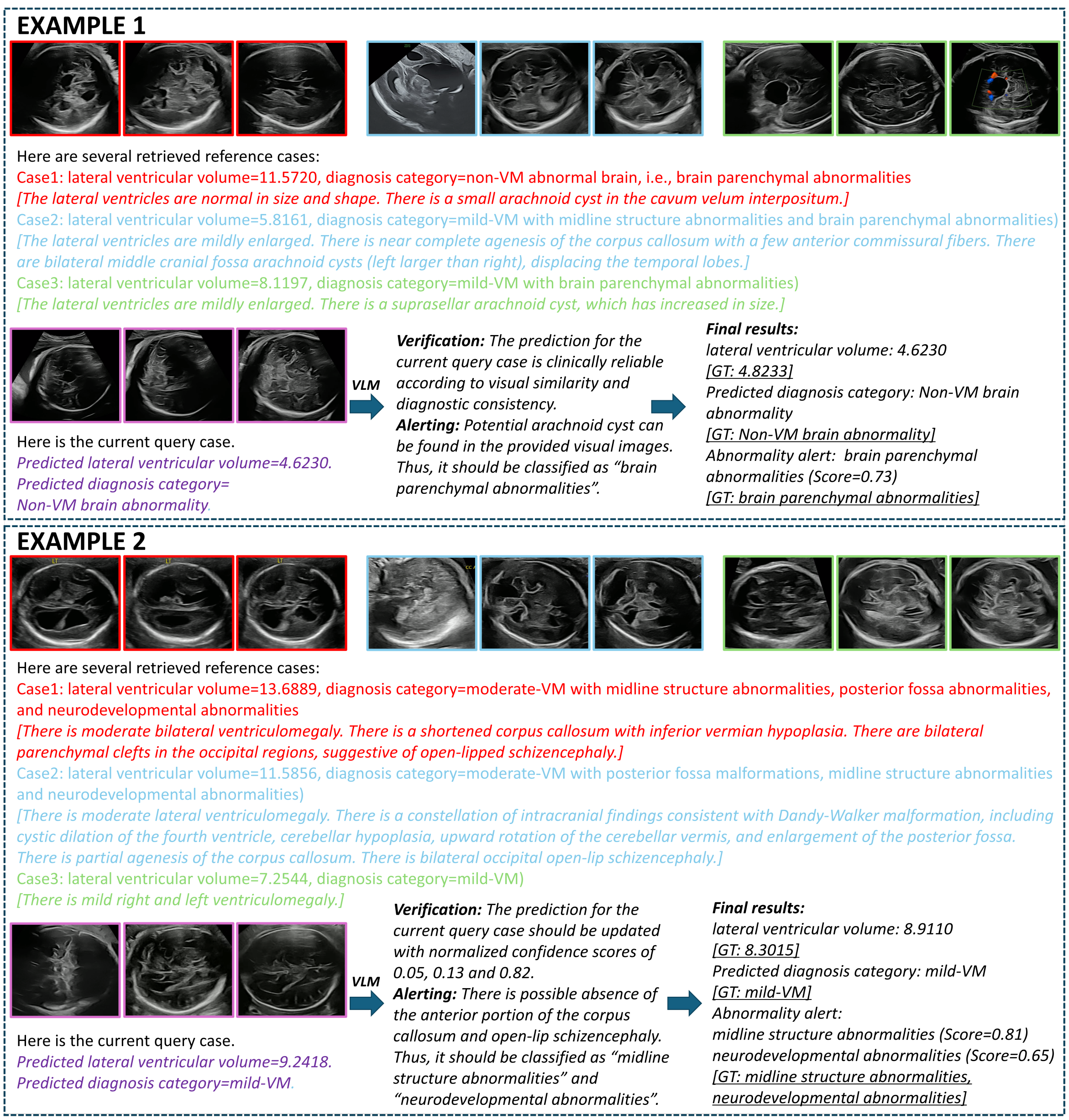}
    \caption{Two typical examples showing the VM verification and multi-abnormality alerting process.}  
    \label{fig:result-vlm}
\end{figure*}

\subsection{Comparison with Different Foundation Model Encoders}

In Table~\ref{tab:result-encoder}, we evaluated the influence of different foundation model-based encoders within the proposed VIFBA framework.
Results show that all encoders achieved strong performance compared with conventional deep models (see Tables~\ref{tab:result-reg} and~\ref{tab:result-cls}), confirming the effectiveness of foundation representations for fetal brain ultrasound video analysis.

Compared with general biomedical encoders (i.e., PMC-CLIP~\cite{lin2023pmc} and BiomedCLIP~\cite{zhang2023biomedclip}), ultrasound-specific foundation models (i.e., USFM~\cite{jiao2024usfm}, EchoCare~\cite{zhang2025fully}, Ultrasound-CLIP~\cite{jin2026ultrasound}, FetalCLIP~\cite{maani2026fetalclip}) generally achieved better performance.
Among all evaluated encoders, FetalCLIP achieved the best error-based regression performance, with the lowest MSE, RMSE and MAE, while maintaining strong correlation and classification performance. 
Figure~\ref{fig:encoder_10} further visualizes the performance variations across 10 independent runs for different encoder variants.
Although some comparisons do not reach statistical significance on specific metrics, significant improvements are observed in the majority of tests (34/60).
This suggests that fetal ultrasound-specific vision-language pretraining provides suitable representations for ventricular volume estimation and VM severity assessment.
The overall results also demonstrate the good extensibility of VIFBA, as it can be flexibly integrated with various foundation encoders while consistently achieving competitive performance.

\subsection{VM Verification and Multi-Abnormality Alerting}

To further evaluate VIFBA's ability to verify VMs and identify non-VM brain abnormalities, we conducted experiments on the multi-abnormality assessment task in Table~\ref{tab:result-vlm}. 
It can be observed that directly extending the classification branch to predict multiple abnormalities yielded limited performance, with the one-stage VIFBA-CLS-S1 achieving an EMR of 0.3200. 
Besides, by first identifying abnormal brains and then predicting the specific abnormality categories, the two-stage VIFBA-CLS-S2 improved the EMR to 0.4100.
These results indicate that explicitly decomposing the task can partially reduce the difficulty of multi-abnormality recognition. 
Nevertheless, conventional supervised classification remains limited in this setting, as fetal brain abnormalities involve diverse categories, may coexist within the same fetus, exhibit substantially imbalanced category distributions with limited samples for rare findings, and affect different anatomical regions across the brain beyond the lateral ventricles.

In contrast, the retrieval-augmented VLM verification strategy substantially improved the overall performance. 
Compared with traditional baselines, all VLMs achieved better evaluation metrics. 
Among them, GPT-5.6 $Sol$ achieved the best overall performance, with a Pre of 0.7207, a Rec of 0.8933, an F1 score of 0.7764, an HL of 0.0440, and an EMR of 0.7500.
GPT-5.6 $Terra$ also showed strong performance, achieving an F1 of 0.7687 and an EMR of 0.7400. These results suggest that VLMs can effectively leverage retrieved visual references and textual descriptions to provide more reliable abnormality-aware screening than task-specific classifiers alone.

Moreover, Figure~\ref{fig:case} shows the qualitative results of ventricular volume regression and multi-abnormality alerting tasks. 
The predicted ventricular volumes closely agree with the ground-truth measurements, and VIFBA successfully identifies potential fetal brain abnormalities in most cases.
Figure~\ref{fig:result-vlm} provides two representative examples illustrating the detailed VLM workflow, including the retrieved reference cases, current query case, verification \& alerting outputs, and the final results.
Each retrieved reference case is accompanied by structured prompts including the \textit{[lateral ventricular volume]} and \textit{[diagnosis category]} information.
 Note that for diagnosis category, we also provide the corresponding description ($[\textit{******}]$, under the structured prompt) in the original report for reference evidence.
Then, given the query case with its predicted lateral ventricular volume and diagnostic category as prompt, our VIFBA will give verification and abnormality-alerting suggestions.
Finally, the predicted volume and category will be retained or updated accordingly, and the potential abnormal type with alert score will also be provided.

Overall, these results demonstrate that retrieval-augmented VLM verification is particularly suitable for multi-abnormality alerting, where diagnostic evidence is sparse, heterogeneous, and often distributed across multiple anatomical regions. By incorporating similar historical cases and clinically relevant textual descriptions, the proposed strategy provides a flexible training-free extension to VIFBA for detecting potential non-VM brain abnormalities beyond VM severity classification.

%% file: conclusion.tex
In this study, we proposed VIFBA, a unified framework for estimating fetal lateral ventricular volume and assessing brain abnormalities directly from ultrasound videos. 
It integrates an ultrasound foundation model with temporal modeling and JEPA-inspired latent predictive learning to enhance video representations.
In addition, MRI-guided cross-modal alignment transfers structural information from MRI to ultrasound representations during training, while requiring only ultrasound at inference.
Finally, a retrieval-augmented, training-free VLM is introduced to verify uncertain predictions and provide auxiliary alerts for potential non-VM brain abnormalities.
Experiments on our large dataset demonstrated that VIFBA substantially outperformed existing representative methods.

Conventionally, ventricular volumetric assessment relies on MRI, whose high cost and limited availability restrict its routine clinical use, particularly in underserved healthcare settings.
To the best of our knowledge, VIFBA is the first exploration to reliably estimate lateral ventricular volume from routinely acquired ultrasound videos, providing a more affordable and widely deployable solution for quantitative assessment.
In addition, our experiments suggest that jointly learning ventricular volume estimation can further improve VM severity classification performance.
This indicates that the regression and classification objectives are not independent, but instead encourage the model to learn shared representations that are transferable across tasks.

Beyond VM assessment, this study further explores the potential of retrieval-augmented VLMs for multi-disease alerting in fetal brain ultrasound.
In clinical practice, except for isolated VM, other fetal brain abnormalities are diverse and often coexist.
However, conventional deep learning models are typically developed for predefined diseases and rely on sufficient disease-specific annotations, limiting their ability to cover rare or underrepresented conditions.
By incorporating retrieved historical cases and their associated diagnostic descriptions as contextual references, our VIFBA enables the plug-and-play integration of off-the-shelf VLMs to compare the query case against multiple abnormal patterns and simultaneously provide alerts for potential non-VM brain abnormalities, without modifying the VLM architecture or requiring additional disease-specific training.
We believe that this strategy provides a flexible pathway for extending disease-specific predictive models toward broader multi-disease screening and more comprehensive fetal brain assessment.

In future work, we will further validate VIFBA on larger multi-center cohorts to assess its generalizability across different ultrasound systems and clinical settings. In addition, we plan to extend the framework toward more comprehensive fetal brain assessment by incorporating additional and fine-grained abnormality categories. Finally, integrating more advanced multimodal foundation models and leveraging larger-scale paired ultrasound–MRI datasets may further enhance the robustness and clinical applicability of the proposed framework.